\documentclass[a4paper]{amsart}
\date{March 23, 2004}

\newtheorem{theorem}{Theorem}[section]
\newtheorem{definition}[theorem]{Definition}

\newtheorem{lemma}[theorem]{Lemma}
\newtheorem{rem}[theorem]{Remark}

\def\const{\mathrm{c}}
\def\cda{c_{\alpha,Z,m,\delta}}
\def\Dd{D^{(\delta)}}
\def\fid{\varphi^{(\delta)}}
\def\Ld{\Lambda^{(\delta)}}
\def\Sd{S^{(\delta)}}

\def\Wd{W^{(\delta)}}
\def\Xd{X^{(\delta)}}

\def\tr{\mathop{\mathrm{tr\,}}\nolimits} 

\newcommand{\cE}{\mathcal{E}}

\newcommand{\gB}{\mathfrak{B}}
\newcommand{\gE}{\mathfrak{E}}
\newcommand{\gH}{\mathfrak{H}}

\newcommand{\gS}{\mathfrak{S}}

\newcommand{\rd}{\mathrm{d}}

\newcommand{\bB}{\mathbf{B}}
\newcommand{\bp}{\mathbf{p}}

\newcommand{\bx}{\mathbf{x}}
\newcommand{\by}{\mathbf{y}}
\newcommand{\bz}{\mathbf{z}}

\newcommand{\cz}{\mathbb{C}} 
\newcommand{\gz}{\mathbb{Z}} 
\newcommand{\nz}{\mathbb{N}} 
\newcommand{\rz}{\mathbb{R}} 

\begin{document}

\author[J.-M. Barbaroux]{Jean-Marie Barbaroux}
\address{D\'epartement de Math\'ematiques\\
  Universit\'e du Sud Toulon-Var\\
  Avenue de l'Universit\'e, BP 132\\
  83957 La Garde Cedex\\
  France} \email{barbarou@univ-tln.fr}

\author[W. Farkas]{Walter Farkas} \address{Swiss Banking Institute\\
  Universit\"at Z\"urich\\
  Plattenstr. 14\\
  8032 Z\"urich\\
  \mbox{Switzerland}} \email{farkas@math.ethz.ch}

\author[B. Helffer]{Bernard Helffer}
\address{D\'epartement de Math\'ematiques\\
  B\^atiment 425\\
  Universit\'e Paris Sud\\
  91405 Orsay C\'edex\\France} \email{Bernard.Helffer@math.u-psud.fr}

\author[H. Siedentop]{Heinz Siedentop} \address{Mathematisches Institut\\
  LMU\\ Theresienstr.
  39\\ 80333 M\"unchen\\ Germany} \email{h.s@lmu.de}

\title{On the Hartree-Fock Equations of the Electron-Positron Field}
\thanks{We thank K.~Yajima for critical reading of a preliminary
  version of the manuscript and D.~Hundertmark for useful and
  enjoyable discussions concerning various aspects of this paper. One
  of us, JMB, thanks B.~Gr\'ebert for useful discussions concerning
  Section \ref{s1}. Financial support of the Minist\`ere de
  l'Education Nationale, de la Recherche et de la Technologie through
  ACI Blanche, of the Bayerisch-Franz\"osisches Hochschulzentrum, of
  the European Union through the IHP network ``Analysis and Quantum'',
  contract HPRN-CT-2002-00277, are gratefully acknowledged. Part of
  the work was done while three of us, JMB, BH, and HS, were visiting
  the Mittag-Leffler Institute, its support is also gratefully
  acknowledged.}

\begin{abstract}
  We study the energy of relativistic electrons and positrons interacting via
  the second quantized Coulomb potential in the field of a nucleus of charge
  $Z$ within the Hartree-Fock approximation. We show that the associated
  functional has a minimizer. In addition, all minimizers are purely
  electronic states, they are projections, and fulfill the no-pair Dirac-Fock
  equations.
\end{abstract}


\maketitle

\section{Introduction}\label{section1}
Heavy atoms should be described by relativistic quantum mechanics.  It is
commonly believed quantum electrodynamics (QED) yields such a description.
Formally the Hamiltonian is given (Bjorken and Drell
\cite[Formula~15.28]{BjorkenDrell1965}) as
\begin{equation}\label{qed}
  \begin{array}{ll}
    H &= \int dx :\psi^*(x) \left[\boldsymbol{\alpha} \cdot
      \left(\frac{1}{i} \nabla - \sqrt{\alpha} \mathbf{A}(x) \right) +
      m\mathbf{\beta} - \frac{\alpha Z}{|\bx|}\right] \psi(x) : \\
    &+    \frac\alpha2 \int dx \int dy
    \frac{:\psi^*(x)\psi(x):\,:\psi^*(y)\psi(y):}{|\bx-\by|} +
    {\frac{1}{8\pi}} \int_{\rz^3} :\bB(\bx)^2 +
    \buildrel{.}\over {\mathbf{A}}(\bx)^2:\,d\bx\;,
  \end{array}
\end{equation}
where the normal ordering denoted by colons is with respect to a given
choice of the one-electron space.  However, it is not clear how this
expression can be self-adjointly realized as a positive operator. To
simplify matters we omit the energy of the transverse radiation field
coupled to the current, i.e., we set $\mathbf{B}=\mathbf{0}$ and
$\mathbf{A} = \mathbf{0}$ in the above expression. We will also
regularize the Coulomb interaction of the electron-positron field by
normal ordering it completely. Both assumptions are simplifications.
The first one can be justified by the physical wisdom that the
presence of self-generated magnetic field is physically known to be
small compared to the relativistic effects in heavy atoms. Moreover,
the inclusion poses serious technical problems that we presently
cannot solve. The second assumption ignores the vacuum polarization
effects which are also small comared to the relativistic effects and
would contribute to the Lamb shift only.

Based on an interesting observation of Chaix, Iracane, and Lions
\cite{ChaixIracane1989,Chaixetal1989}, Bach et al.
\cite{Bachetal1999,Bachetal1998} showed positivity for the corresponding
quadratic form without any constraints on the charge of the state, if the
one-electron subspace is appropriately (Furry picture) chosen. In particular
they showed that the vacuum state (particle number equal to zero) has energy
zero.

In order to do so, they proved that the positivity of $H$ on generalized
Hartree-Fock states (quasifree states with finite free kinetic energy) is
equivalent to the positivity of the Hartree-Fock functional (see the
Definition given in \eqref{def1}), a functional on density matrices $\gamma$,
where the charge is given as $\tr\gamma$, which thus is unrestricted.

For describing atoms one needs, however, to restrict to states with prescribed
charge $q$. In order to implement this problem, we subtract from the energy
the rest mass $m$, which -- as we will see -- will allow us to relax the
constraint $\tr\gamma=q$ to $\tr\gamma\leq q$.

Given an electron space by the positive spectral subspace of a Dirac operator
with a mean-field potential, we show that there exists a minimizer of the
associated Hartree-Fock functional in a suitable set, if $q\leq Z$
(Theorem~\ref{sec:existence-cor}).  Moreover, if $q$ is a positive integer,
all minimizers are projections of maximal rank. In particular, minimizers are
purely electronic and are projections (Theorem~\ref{sec:min-are-proj-1}).
Finally, we show that the eigenfunctions of any minimizer $\gamma$ fulfill the
no-pair Dirac-Fock equations with the $q$ first (positive) eigenvalues
(Theorem~\ref{th:proj}).

According to Mittleman \cite{Mittleman1981}, the most stable, i.e.,
highest ground state energy is the physical ground state. Thus one
should maximize among the allowed one-electron subspaces, yielding a
max-min variational problem. The resulting Euler equation should, on a
heuristic level, give the Dirac-Fock equations which were treated by
Esteban, S\'er\'e, and Paturel
\cite{EstebanSere1999,Paturel2000,EstebanSere2001}. There are
indications that this latter question might be answered affirmatively
only under additional hypotheses: In a recent work
\cite{Barbarouxetal2004} an atom with total charge $q-Z\leq 0$ is
considered. If the ground state of the noninteracting problem with
$N=q$ electrons corresponds to a closed shell, then for small
interaction maximizing over one-electron subspace yields the
Dirac-Fock equations in the non-relativistic limit. However, if the
noninteracting problem corresponds to an open shell, then, in the same
limit, the max-min procedure does not yield a solution of the
Dirac-Fock equations with self-consistent projector as considered by
Esteban and S\'er\'e \cite{EstebanSere1999}. While this result is
perturbative, it indicates on the one hand side that the Dirac-Fock
equations and the Mittleman principle might agree in the case of filled
shells whereas in the unfilled shell case it might give different
results which raises the question which procedure is physically
relevant, a problem that we have to leave open at this point.

We add a short guide through the paper for the orientation of the
reader: Section~\ref{section2} contains some basic material. We define
the set of density matrices that will be allowed. There will be two
types of density matrices, the charge density matrices $\gamma$ for
the electron-positron field and the density matrices $\delta$ giving
the screening of the one-particle Dirac operator that defines the
electron subspace. In addition this section contains some basic
estimates on the direct and exchange energy. Section \ref{s1} contains
the actual minimization. We first show that the elimination of
positrons lowers the energy (Lemma \ref{diff-proj}); next we show that
the density matrices can be restricted to finite rank (Lemma
\ref{sec:reduct-unren-dens}), and the minimization under the
constraint $\tr \gamma \leq q\leq Z$ gives a minimizer -- if existing
-- with charge equal to $q$ (Lemma \ref{t1}). This allows us to show
the existence of minimizers (Theorem \ref{sec:existence-cor}). In
Section \ref{s7} we investigate the minimizers.  They turn out to be
projections that fulfill the no-pair Dirac-Fock equations (Theorem
\ref{th:proj}).  In the last section we give an outlook with respect
to the above mentioned program of Mittleman.  We derive the
corresponding Euler equation (Theorem \ref{wbr}).  However, we are not
able to show that there is a maximizer.

\section{Definition of the Problem}\label{section2}

A single relativistic electron or positron in the field of a nucleus of charge
$Z$ can be described by the Coulomb-Dirac operator
\begin{equation}
  \label{eq:1} D_Z := \boldsymbol{\alpha}\cdot \frac1i\nabla +
  m\beta - \alpha \frac{Z}{| \bx |}\;,
\end{equation}
where $\alpha$ is the Sommerfeld fine structure constant.

The operator $D_Z$ is self-adjointly realized in $\gH:= L^2(\rz^3)\otimes
\cz^4$ and essentially self-adjoint on $C_0^\infty(\rz^3\setminus\{0\})\otimes
\cz^4$, if $\alpha Z\in(-\sqrt3/2,\sqrt3/2)$.  Here, we will even assume
 \begin{equation}\label{eq:hyp-alphaZ}
 \alpha Z\in [0,\sqrt3/2)\quad\mbox{and}\quad
 \alpha \geq 0 \;,
 \end{equation}
 throughout the paper. The domain of $D_Z$ is $H^1(G)$ where
 $$
 G:=\rz^3\times\{1,2,3,4\}\;.
 $$
 (Landgren and Rejto \cite{LandgrenRejto1979}, Theorem~2.1). For $Z=0$,
 $D_0$ is just the free Dirac operator.

 In the present paper, we consider a larger class of Dirac operators, namely
 Dirac-Fock operators. They are Hamiltonians for a relativistic particle in a
 mean field created by other particles. For that purpose, we will consider
 operators with an additional mean field potential
\begin{equation}\label{D-F-potential}
  \Wd = \fid { -\Xd}\;,
\end{equation}
where $\fid$ and $\Xd$ will be defined below.

For $p\in [1,\infty)$, we denote by $\gS_p(\gH)=\{ A\in\mathcal{B}(\gH) \ |\
\tr|A|^p <\infty\}$, and by $\gS_\infty(\gH)$ the space of compact operators
on $\gH$.
\begin{definition}\label{def:F}
  We denote by $F$ the Banach space of all self-adjoint operators $\delta$ on
  $\gH$ such that the norm
  $\|\delta\|_F:=\tr(\,|D_0|^{1/2}\,|\delta|\,|D_0|^{1/2}\,)$ is finite.
\end{definition}
For a given element $\delta\in F$, we denote by $(\lambda_n)$ the sequence of
its eigenvalues and by $(\xi_n)$ a corresponding orthonormal basis of
eigenvectors; the associated integral kernel $\delta(x,y)$ is
\begin{eqnarray}
  \label{eq:kern}\delta(x,y)&:=&\sum_n \lambda_n
  \xi_n(x)\overline{\xi_n(y)}\;.
\end{eqnarray}
(It is convenient to introduce the notation $x=(\bx ,s)$ for an element of $G$
and ${\mathrm d}x$ for the product of the Lebesgue measure ${\mathrm d}{\bf
  x}$ on $\rz^3$ with the counting measure in $\{1,\,2,\,3,\,4\}$.)
Associated with $\delta$ is its one-particle density
\begin{equation}
  \label{density}
  \rho_\delta(\bx) := \sum_{s=1}^4\sum_n \lambda_n |\xi_n(x)|^2\;,
\end{equation}
its electric potential
 \begin{equation}\label{direct}
  \fid(\bx) =  \int_{\rz^3}  \frac{\rho_\delta(\by)}{|\bx -\by|}\rd\by\;,
 \end{equation}
 and its exchange operator $\Xd$
\begin{equation}\label{exchange}
  \psi \mapsto   \int\frac{{\delta(x,y)}\psi(y)}{| \bx - \by|} \rd y\;.
\end{equation}
The difference of these two operators is the mean field potential $\Wd$
defined in \eqref{D-F-potential}.  Next, we define for the given $\delta$ the
Coulomb-Dirac operator associated to $\delta$ as
\begin{equation}\label{D-F-operator}
    \Dd := D_Z + \alpha \Wd\;.
\end{equation}

As shown in the Appendix \ref{section3} (Lemmata~\ref{l3tierce} and
\ref{lem:Xsmallphi}), the operator $\Wd$ is bounded implying that $\Dd$ is
self-adjoint with the same domain as the Coulomb-Dirac operator $D_Z$ which
for $\alpha Z \in [0,\sqrt3/2)$ is identical with the domain of $D_0$.
Moreover (see Lemma~\ref{compactness}), $\Wd$ is relatively compact with
respect to $D_0$ which implies
$$
\sigma_\mathrm{ess}(\Dd) = \sigma_\mathrm{ess}(D_0)=(-\infty,
-m]\cup[m,\infty).
$$
Finally, since $|\Dd|\geq \cda|D_0|$ by definition of $\cda$ (Equation
\eqref{eq:moduli}) and since $|D_0|>0$, the operator $\Dd$ has a bounded
inverse, as soon as $\cda>0$.

The one-electron states are vectors in $\gH_+=\Lambda_+\gH$ where $\Lambda_+$
is an orthogonal projection on $\gH$, whereas one-positron states are charge
conjugated states in $\Lambda_-\gH$, where $\Lambda_-:=1-\Lambda_+$
(\cite{Thaller1992}). We will take $\Lambda_+$ to be the projection $\Ld_+$
onto the positive spectral subspace of the Dirac-Fock operator $\Dd$, i.e.,
\begin{equation}
  \label{eq:2}
  \Lambda_+:=\Ld_+ :=\chi_{[0,\infty)}(\Dd)\;,
\end{equation}
where $\chi_I$ denotes the characteristic function of the set $I$.  Thus, the
choice of $\delta$ fixes the definition of the spaces of electrons and
positrons.

The Coulomb scalar product is
\begin{equation}
  \label{coulomb}
  D[\rho,\sigma]:=\frac12 \int_{\rz^3}\rd\bx \int_{\rz^3} \rd\by
  \frac{\overline{\rho(\bx)}\sigma(\by)}{| \bx - \by|}\;.
\end{equation}
The exchange scalar product for $\gamma,\gamma'\in F$ is
\begin{equation}
  \label{eq:austausch}
  E[\gamma,\gamma'] :=\frac12\int\rd x\int\rd y
  \frac{\overline{\gamma(x,y)}\gamma'(x,y)}{|\bx-\by|}\;.
\end{equation}

\begin{lemma}
  \label{sec:definition-problem}
  Assume that $\gamma,\gamma'\in F$. Then
  \begin{eqnarray}
    \label{eq:q1}
    |D[\rho_\gamma,\rho_{\gamma'}]| &\leq& \frac\pi4
     \|\gamma\|_1\tr(\sqrt{-\Delta}|\gamma'|)\;,\\
    \label{eq:3}
    |E[\gamma,\gamma']|&\leq&
    D[\rho_{|\gamma|},\rho_{|\gamma'|}]\;.
  \end{eqnarray}
\end{lemma}
\begin{proof}
  Expanding $\gamma$ and $\gamma'$ in their respective bases of eigenfunctions
  (see \eqref{eq:kern}), we get by the Cauchy-Schwarz inequality
  \begin{multline}
    \label{wellq12}
    \left|\int\int \frac{\overline{\gamma(x,y)}\gamma'(x,y)}{| \bx-\by|}\rd
      x\, \rd y\right| = \left|\int \int \sum_\mu \lambda_\mu \sum_\nu
      \lambda'_\nu
      \frac{\overline{\xi_\mu(x)}\xi_\mu(y)\xi'_\nu(x)\overline{\xi'_\nu(y)}}
      {|\bx-\by|}\rd x \rd y \right|\\
    \leq \int \int \frac{\sum_\mu|\lambda_\mu| |\xi_\mu(x)|^2
      \sum_\nu|\lambda'_\nu||\xi'_\nu(y)|^2} {|\bx-\by|}\rd x \rd y =
    \int_{\rz^3} \int_{\rz^3}
    \frac{\rho_{|\gamma|}(\bx)\rho_{|\gamma'|}(\by)}{| \bx - \by|}\rd \bx\,
    \rd \by \;,
  \end{multline}
  which shows that it suffices to prove \eqref{eq:q1}.

  To this end we remark that by Kato's inequality
  \begin{equation}
    \label{eq:4}
    \int dx|\xi_\mu(x)|^2\int dy\frac{|\xi'_\nu(y)|^2} {|\bx-\by|}\leq
    \frac\pi2(\xi'_\nu,|\nabla|\xi'_\nu)\;.
  \end{equation}
  The claimed inequality follows now by multiplication with
  $|\lambda_\mu\lambda'_\nu|$ and summation over $\mu$ and $\nu$.
\end{proof}
We will also need the following result of Bach et al (\cite{Bachetal1999},
Equation (30))~:
\begin{lemma}
  \label{sec:definition-problem-1}
  If $\gamma\in F$, then
  $$E[\gamma,\gamma]\leq\frac\pi4\,
  \tr\left(\gamma^*\,\sqrt{-\Delta}\,\gamma\right)\;.$$
\end{lemma}
Given an operator $A$ on the Hilbert space $\gH$ the symbols
$A_{++}=\Ld_{+}A\Ld_+$, $A_{+-}=\Ld_{+}A\Ld_-$, $A_{-+}=\Ld_-A\Ld_+$, and
$A_{--}=\Ld_{-}A\Ld_-$ denote the matrix elements of the decomposition of $A$
with respect to the splitting of the Hilbert space given by $\Ld_+$ and
$\Ld_-$ (We assume that the operators $\Ld_\pm$ leave the domain of $A$
invariant).

Relativistic electrons and positrons are described by one-particle charge
density matrices $\gamma$ with certain additional properties reflecting the
charge of the particle, and the fact that they are Fermions and thus obey the
Pauli principle.
\begin{definition}
  Given $\delta\in F$ and $q\in\rz_+$ we define the following sets of
  one-particle diagonal charge density matrices:
  \begin{eqnarray}
    \label{set:1pdm}
    \Sd &:=& \{ \gamma\in F\ |\ -\Ld_-
    \leq \gamma \leq \Ld_+, \Ld_- \gamma \Ld_+=0\}\;,\\
    \label{set:1pdmq}
    \Sd_q &:=& \{ \gamma\in \Sd\ |\ 0\leq \tr\gamma\leq q\}\;,\\
    \label{set:1pdmdq}
    \Sd_{\partial q}&:=& \{\gamma\in \Sd\ |\ \tr\gamma=q\}\;.
  \end{eqnarray}
\end{definition}
We note that all sets are closed subsets of $F$. Furthermore, the first two
are convex. Note also that for $\gamma\in\Sd$ we have $\gamma_{++}\geq 0$ and
$\gamma_{--}\leq 0$ which follows directly from the definition. We also
observe that~:
\begin{eqnarray}
  \label{f1}
  \gamma_{++}^2 &\leq& \gamma_{++}\;,\\
  \label{f2}
  \gamma_{--}^2 &\leq& -\gamma_{--} \;,
\end{eqnarray}
which permits to get (in the case of diagonal density matrices)
 \begin{eqnarray}\label{eq:22}
  \tr(|\Dd| |\gamma|)=\tr(\Dd\gamma) \geq \tr(|\Dd|\gamma^2)\;.
 \end{eqnarray}
 The elements of $\Sd$ are the one-particle (renormalized) charge density
 matrices $\gamma$ of the electron-positron field. The trace is its total
 charge.  Since we are interested in describing atoms we want to fix the
 charge of the electron-positron field to be $q$ and minimize the energy over
 the set $\Sd_{\partial q}$.  For technical reasons we will also use $\Sd_q$.

 We wish to point out that he derivation of the variational spaces of
 one-particle charge density matrices as done in \cite{Bachetal1999} does not
 give the extra condition $\Ld_-\gamma\Ld_+ = 0$ that appears in the
 definition of $\Sd$. A formal calculation shows that if we do not assume that
 the one-particle density matrices have off diagonal terms equal zero, then in
 most cases, Lemma~\ref{diff-proj} and Theorem \ref{sec:min-are-proj-1} and
 \ref{th:proj} do not hold. In particular, minimizers will contain
 electron-positron pairs.

 The projections $\Lambda_-^{(\delta)}$ can be physically interpreted as the
 one-particle density matrix of the Dirac sea.  In particular, $\Ld_-$ is the
 Dirac sea under the influence of a nucleus of charge $eZ$ and an
 electron-positron distribution given by $\delta$.

 For later purposes we also introduce (unrenormalized) density matrices as
 $$
 \Gamma:=\Ld_- + \gamma\;,
 $$
 representing all electrons including those of the Dirac sea.

 The unrenormalized density matrices are nonnegative expressing the fact that
 positrons occur in this picture only as `holes' in the Dirac sea.

 The energy of a system of electrons and positrons in Hartree-Fock
 approximation is given by the functional
 \begin{equation} \label{def1}
  \cE :
  \begin{array}{lll}
    \Sd & \rightarrow & \rz \\
    \gamma &\mapsto &\tr(D_Z\gamma) +    \alpha
    Q[\gamma,\gamma]\;,
  \end{array}
 \end{equation}
 where
\begin{equation} \label{def-Q}
 Q :
  \begin{array}{lll}
    \Sd\times\Sd & \rightarrow & \rz \\
    (\gamma,\gamma') &\mapsto &  D[\rho_\gamma,\rho_{\gamma'}]
     - E[\gamma,\gamma'] \;.
  \end{array}
\end{equation}

As explained above we are primarily interested in the infimum of
$\cE|_{\Sd_{\partial q}}$; for technical reasons we will also consider, for
$\mu\in\rz$,
 \begin{equation} \label{def1m}
  \cE_\mu :
  \begin{array}{lll}
    \Sd & \rightarrow & \rz \\
    \gamma &\mapsto &\cE(\gamma) - \mu\tr\gamma
  \end{array}
 \end{equation}

\begin{lemma}
  \label{welldefined}
  For any $\mu\in\rz$, the energy functional $\cE_\mu$ is well defined and
  continuous in the $\|\cdot\|_F$ norm.
\end{lemma}
\begin{proof}
  The lemma is an immediate consequence of the definition of the norm, Lemma
  \ref{sec:definition-problem-1} together with the fact that
  $\gamma^2\leq|\gamma|$ and Inequalities \eqref{eq:q1} and \eqref{eq:3}.
\end{proof}

\section{Minimization of the Energy\label{s1}}

\subsection{Reduction to Electrons}
\label{sec:reduction-electrons}

\begin{lemma}
\label{diff-proj} Assume $q>0$, $\gamma\in\Sd_{\partial q}$,
$0\leq\delta\in F$. Moreover, assume
$$
\cda > \pi\alpha(1/4+\max\{\tr\delta,q\})\;.
$$
Then there exists a nonnegative $\gamma_e\in\Sd_{\partial q}$ and
$R\in\Sd_{\partial 0}$ such that $\gamma=\gamma_e+R$ and $\cE(\gamma_e)\leq
\cE(\gamma)$.  In addition, equality can only occur if $0\leq\gamma$.
\end{lemma}
Physically speaking this lemma says that it is favorable to have no positrons
in the system and to restrict the minimization to electron states.
\begin{proof}

  Using the spectral decomposition of $\gamma$, one can easily construct
  $\gamma_e$ and $R$ such that $\gamma_e\in\Sd_{\partial q}$,
  $R\in\Sd_{\partial 0}$ and $\gamma=\gamma_e+R$. We also note that we can
  pick $R\neq0$, if $\gamma\not\geq0$, and that we can pick $R=0$, if
  $\gamma\geq0$. We have
  \begin{equation}
    \label{eq:proof-diff-proj1}
    \begin{split}
      \cE(\gamma_e) - \cE(\gamma)
      &= -\tr( D_Z R) -\alpha Q[R,R]- 2\alpha \Re Q[R, \gamma_e]\\
      & = -\tr(\Dd R) - \alpha Q[R,R] - 2 \alpha \Re
      Q[R,\gamma_e] + 2\alpha \Re Q[R, \delta]\\
      &\leq-\tr(\Dd R) + \alpha E[R,R] - 2\alpha \Re Q[R,\gamma_e] + 2\alpha
      \Re Q[R, \delta]\;,
    \end{split}
  \end{equation}
  where in the last inequality, we used positivity of $D[\rho_R, \rho_R]$.

  Let $R_+$ and $R_-$ be respectively the positive and negative part of $R$,
  i.e.
  $$
  R = R_+ - R_- \mbox{ with } R_\pm \geq 0\;.
  $$
  Using Lemma~\ref{lem:Xsmallphi} and the positivity of $\gamma_e$ yields
  $$
  \Re Q[R_+, \gamma_e]\geq 0 \mbox{ and thus }-\Re Q[R,\gamma_e] \leq \Re
  Q[R_-,\gamma_e]\;.
  $$
  Similarly, we have $\Re Q[R, \delta] \leq \Re Q[R_+, \delta]$.
  Therefore, from Inequality \eqref{eq:proof-diff-proj1} we obtain
 \begin{equation}\label{eq:proof-diff-proj1-a}
    \begin{split}
      \cE(\gamma_e) - \cE(\gamma) &\leq-\tr(\Dd R) + \alpha E[R,R] + 2\alpha
      \Re Q[R_-,\gamma_e] + 2\alpha \Re Q[R_+, \delta]\;.
    \end{split}
  \end{equation}
  From Lemma \ref{sec:definition-problem-1} and the definition of $\cda$ in
  \eqref{eq:moduli}, we get
\begin{equation}\label{eq:proof-diff-proj1-b}
  E[R,R] \leq \frac\pi4 \tr(|D_0| R^2) \leq
  \frac{\pi}{4\cda} \tr( |\Dd| R^2) \leq
  \frac{\pi}{4\cda} \tr( \Dd R)\;,
\end{equation}
where the last inequality is a consequence of \eqref{eq:22}.

Using \eqref{eq:3}, \eqref{eq:q1}, and, finally, the definition of $\cda$ in
\eqref{eq:moduli} we have
  \begin{equation}
    \begin{split}
      \Re Q[R_-,\gamma_e] &= D[\rho_{R_-},\rho_{\gamma_e}]- \Re
      E[R_-,\gamma_e] \leq 2 D[\rho_{R_-},\rho_{\gamma_e}] \\
      &\leq \frac\pi2 \tr\gamma_e \tr(|p|R_-)\leq \frac{\pi
        \tr\gamma_e}{2\cda} \tr(|\Dd|R_-)\;.
      \label{eq:electron2}
    \end{split}
  \end{equation}

  With exactly the same arguments as above, exchanging $R_-$ by $R_+$ and
  $\gamma_e$ by $\delta$ we get
  \begin{equation}
    \Re Q[R_+,\delta] \leq \frac{\pi \tr\delta}{2\cda}\tr (|\Dd| R_+).
    \label{eq:added}
  \end{equation}
  Since $\tr(|D^{(\delta)}| |R|) = \tr(D^{(\delta)} R)$, we have, using
  \eqref{eq:proof-diff-proj1-a}-\eqref{eq:added}
  \begin{equation}\label{punch}
    \cE(\gamma_e) - \cE(\gamma)
    \leq \left(-1 +\frac{\pi\alpha}{4\cda}
+\frac{\pi\alpha\max\{\tr\delta,q\}}\cda
    \right)\tr(\Dd R)\leq 0\;,
  \end{equation}
  under the hypothesis of the theorem. In addition we note that the last
  inequality is strict unless $R=0$ or the prefactor vanishes.
\end{proof}

\subsection{Reduction to Density Matrices with Finite Spectrum}
\label{s1.1}

\begin{lemma}
  Assume $0\leq q$, $0\leq\delta\in F$ and $0\leq \gamma \in
  \Sd_{\partial q}$. Then there exists a sequence of finite rank
  density matrices $\gamma_K\in \Sd_{\partial q}$ such that
  $\|\gamma_K-\gamma\|_F\to0$ as $K\to\infty$.
\end{lemma}

\begin{proof}
Let $(\xi_k)_{k\in \nz}$ be a complete set of eigenfunctions in
$H^{1}(G)$ of $\gamma$ associated with the eigenvalues
$\lambda_k$. Since $\gamma\geq 0$, we have
$\gamma=\Ld_+\gamma\Ld_+$ and, for all $k$, we have
$\xi_k\in\Ld_+\gH$.

Assume that $\gamma$ is not already of finite rank. Then, since
$\gamma$ is trace class, there exist infinitely many eigenvalues
of $\gamma$ in $(0,1)$. Let us pick $\lambda_{n_0}\in (0,1)$, one
of these eigenvalues.

Set $\epsilon_K:= q- \sum_{k=1}^K \lambda_k$. Then $\epsilon_K$ is
a nonnegative monotone decreasing sequence tending to zero. Define
$\gamma_K := \sum_{k=1}^K \lambda_k |\xi_k\rangle\langle\xi_k| +
\epsilon_K|\xi_{n_0}\rangle\langle\xi_{n_0}|$.

We assume that $K$ is chosen sufficiently large so that $K\geq
n_0$ and $\lambda_{n_0}+\epsilon_K < 1$. Obviously each $\gamma_K$
is nonnegative, belongs to $\Sd_{\partial q}$ and has finite rank.

We now show that $\gamma_K\to\gamma$ in F-norm as $K\to\infty$. We
have
 $$
 \gamma-\gamma_K = \sum_{k=K+1}^\infty \lambda_k
 |\xi_k\rangle\langle\xi_k| - \epsilon_K|\xi_{n_0}\rangle\langle\xi_{n_0}|\;.
 $$
Thus,
  \begin{equation}
    \label{202}
    \|\gamma-\gamma_K\|_F\leq \sum_{k=K+1}^\infty \lambda_k
 \tr(|D_0||\xi_k\rangle\langle\xi_k|) + \epsilon_K (\xi_{n_0},|D_0|\xi_{n_0})\;.
  \end{equation}
The first in the right hand side tends to zero, since
$|D_0||\gamma|\in\gS_1(\gH)$, and the second tends to zero, since
$\epsilon_K\to 0$.
\end{proof}
The following is an immediate consequence of the continuity of
$\cE$ in the $F$-norm and the preceding density result.
\begin{lemma}
  \label{sec:reduct-unren-dens}
  Assume that $q>0$. Then
  $$\inf\{\cE_\mu(\gamma)\; |\; 0\leq\gamma\in \Sd_{\partial q}\} =
  \inf\{\cE_\mu(\gamma)\; |\; 0\leq\gamma \in \Sd_{\partial q},
  \mathrm{rank}(\gamma)<\infty \}.$$
\end{lemma}

\subsection{Reduction to Projection} \label{sec:reduct-proj}

Following an argument of Bach \cite{Bach1992}, we get
\begin{lemma}
  \label{sec:reduct-proj-1}
Assume $q\in\nz$, $\delta\in F$, $0\leq\gamma\in \Sd_{\partial q}$
with finite rank. Then there exists a projection
$P\in\Sd_{\partial q}$ such that
  \begin{equation}\label{inrepro}
\cE(P)\leq\cE(\gamma)\;.
\end{equation}
Equality in \eqref{inrepro} holds only if $\gamma$ is already a
projection.
\end{lemma}
\begin{proof}
Suppose that $\gamma$ is not a projection. Then $\gamma$ has an
eigenvalue $\lambda\in(0,1)$; we denote a corresponding normalized
eigenvector by $u$ and observe that it is in $H^1(G)$. Since
$\tr(\gamma)\in \nz$, there exists at least a second eigenvalue
$\mu\in(0,1)$; we denote a corresponding normalized eigenvector by
$v$. We set
  \begin{equation}
    \label{eq:9}
    \tilde\gamma:=\gamma + \epsilon S\;,
  \end{equation}
where $S:=|u\rangle\langle u|-|v\rangle\langle v|$. We get
  \begin{equation}
    \label{eq:10}
    \cE(\tilde\gamma)-\cE(\gamma)=
    \epsilon(\tr(D_ZS)+2\Re Q[\gamma,S]) +\epsilon^2 Q[S,S]\;.
  \end{equation}
By explicit computation and use of the Cauchy-Schwarz inequality,
one can  show that~:
 \begin{equation}\label{QSS}
Q[S,S]<0\;,\; \mbox{ if } S \mbox{ is a difference of two
orthogonal rank one projections}.
\end{equation}
Now -- depending on the sign of
the coefficient linear in $\epsilon$ -- we lower the energy by
increasing or decreasing $\epsilon$ from zero, until one of the
constraints $ 0\leq \lambda + \epsilon, \mu-\epsilon\leq1$ forbids
any further increase or decrease of $\epsilon$.  This process
leaves all the eigenvalues of $\gamma$ unchanged except for $\mu$
and $\lambda$, one of which becomes either $0$ or $1$.

Since there are only finitely many eigenvalues of $\gamma$
strictly between zero and one (even if they are counted according
to their multiplicity), iterating this process eliminates all
eigenvalues that are strictly between $0$ and $1$, i.e., we have
found a density matrix $P$ such that $P^2=P$.
\end{proof}
\begin{rem}\label{rem:charge}
Following the same method in the case of $q\geq 0$, not
necessarily integer, we can show that, given $0\leq\gamma\in
S_{\partial q}^{(\delta)}$, there exists $\tilde P$ equals a
projection plus a rank one operator such that $\tr \tilde{P} =
\tr\gamma$, $\tilde{P}\geq 0$ and $\cE(\tilde{P})\leq
\cE(\gamma)$, with equality only if $\gamma$ is already a
projection plus a rank one operator.
\end{rem}

\subsection{Lower Bound on the Energy}
\label{sec:minimization-energy}

In this subsection, we show that for sufficiently small fine
structure constant $\alpha$ and atomic number $Z$, the energy is
bounded from below.
\begin{theorem}
  \label{sec:minimization-energy-2}
Assume $0\leq\delta\in F$, and
$\cda\geq\pi\alpha(1/4+\max\{\tr\delta,q\}) > 0$. Then, for all
$\gamma\in \Sd_q$, $\cE(\gamma)\geq 0$.
\end{theorem}
\begin{proof}
By Lemma \ref{diff-proj} we need to consider only positive
$\gamma$'s. In this case, \eqref{eq:3} implies $Q[\gamma,\gamma]
\geq 0$. Now, for
  $f\in \Ld_+\gH$, using the definition \eqref{eq:moduli} of
  $\cda$, Inequality \eqref{eq:q1} and the positivity
  of $E^{(\delta)}$, we obtain
  \[
    (f, D_Zf) = (f,|\Dd|f) -\alpha(f,\Wd f)\geq \cda(f,|p|f) -
    \frac\pi4\alpha \tr\!\delta(f,|p|f)\geq 0\;.
  \]
  Thus, under our hypotheses,  $ \cE(\gamma) \geq \tr(D_Z\gamma) \geq
  0$.
\end{proof}

\subsection{Reduction to Density Matrices of Maximal Charge}
\label{sec:reduct-dens-matr}
\begin{lemma}\label{t1}
Assume  $0\leq q$, $0\leq\delta\in F$ and suppose that, for all
$0\leq p<q$ and all $0\leq\gamma\in \Sd_p$, the operator
$\Ld_+D^{(\gamma)}\Ld_+$ has infinitely many eigenvalues in
$(0,m)$. Then
  $$
  \inf\left\{\cE_m(\gamma)|\gamma\in \Sd_{\partial q},\gamma\geq 0\right\} =
  \inf\left\{\cE_m(\gamma)|\gamma\in \Sd_q,\gamma\geq 0\right\}\;.
  $$
If in addition $0\leq\tilde\gamma$ is a minimizer of $\cE_m$ in $\Sd_q$, it
  follows that $\tr \tilde\gamma=q$.
\end{lemma}
\begin{proof}
That the left side bounds the right side from above is obvious. To
prove the converse inequality, we assume $0\leq\gamma\in \Sd_q$,
with $\tr \gamma < q$. By Lemma~\ref{sec:reduct-proj-1} and
Remark~\ref{rem:charge} we can assume that $\gamma$ is a
projection plus a rank one operator; in particular its range is
finite dimensional. Since by assumption the discrete spectral
subspace $\mathfrak{X}$ of $\Ld_+D^{(\gamma)}\Ld_+$ is infinite
dimensional we can find $u\in \gH_+\cap \mathfrak{X}\cap
\gamma(\gH)^\perp$ with $0< \|u\|\leq 1$ and define
$\tilde\gamma:=\gamma+A$ with $A:=|u><u|$. We then get
  $$
    \cE_m(\gamma+A)-\cE_m(\gamma)=\tr(D_ZA)+ 2\Re Q(\gamma,A)- m
    \|u\|^2
    =\langle u,(D^{(\gamma)}-m)u\rangle<0\;.
  $$
Thus, this construction yields a density matrix $\tilde\gamma$
with strictly smaller energy and a trace that can be made bigger
by $\min\{1,q-\tr\gamma\}$. Iteration of the construction yields
the desired result.  This proves both claims.
\end{proof}

\subsection{Existence of a Minimizer}
\label{s1.2}

We wish to show the existence of a minimizer by weak lower
semi-continuity of the functional on a minimizing sequence and weak
compactness. However, we are faced with the problem that we are
minimizing on charge density matrices and the fact that the Coulomb
potential is not relatively compact with respect to the relativistic
energy $|p|$.  The first problem has been addressed by Solovej
\cite{Solovej1991} in the non-relativistic context. To handle the
second problem we will decompose the one-particle part of the energy.

\begin{theorem}\label{existence}
Assume $0<\cda$, $0\leq\delta\in F$ and $q\in\nz$. Furthermore,
suppose that $\cE_m$ is bounded below on $\Sd_{\partial q}$ and
  $$
   \inf\{\cE_m(\gamma)\; |\; \gamma\in \Sd_q\}=\inf\{\cE_m(\gamma)\;
   |\; \gamma\in\Sd_{\partial q},\gamma^2=\gamma\}\;.
  $$
Then the energy functional $\cE_m|_{\Sd_q}$ has a minimizer.
\end{theorem}
\begin{proof}
  Let $\gamma_n$ be a minimizing sequence of orthogonal projections in
  $\Sd_{\partial q}$.

\textbf{Step 1. Weak Limit of the Minimizing Sequence:} Since
$\cE_{m}(\gamma_n)$ and $\fid$ are bounded (Lemma \ref{l3tierce})
(and thus also $\Xd$, by Lemma~\ref{lem:Xsmallphi}, and $\Wd$)
there exists a constant $C$ such that, for any $p\geq 1$
  \begin{multline*}
  C\geq \cE_m(\gamma_n) + \alpha q\|\Wd\| \geq \tr(|\Dd|\gamma_n) \geq \cda
  \tr(|D_0|\gamma_n)=\cda\|\gamma_n\|_F \\
  \geq
  \cda\||D_0|^{1/2}\gamma_n|D_0|^{1/2}\|_p\; .
\end{multline*}
Thus, according to Banach and Alaoglu, there exists,  for all
$p>1$,  $\gamma_\infty$ with
$\||D_0|^{1/2}\gamma_\infty|D_0|^{1/2}\|_p<\infty$ and a
subsequence of $\gamma_n$ -- also denoted by $\gamma_n$ -- such
that for all $B$ with $\|B\|_q<\infty$,
\begin{equation}
    \label{eq:weak}
    \tr(B|D_0|^{1/2}\gamma_n|D_0|^{1/2})\to
    \tr(B|D_0|^{1/2}\gamma_\infty|D_0|^{1/2})\;,
\end{equation}
where $1/p+1/q=1$.

We denote $\|\gamma_n\|_{F,p} := \||D_0|^{1/2}\gamma_n|D_0|^{1/2}
\|_p$. Given $p_1 \geq p_2 \geq 1$, it is always possible to
extract a subsequence of $\gamma_n$ -- again denoted by $\gamma_n$
--, such that it converges weakly in $(F, \|.\|_{F,p_1})$ and $(F,
\|.\|_{F,p_2})$. Denote the weak limits respectively by
$\gamma^{(p_1)}_\infty$ and $\gamma^{(p_2)}_\infty$. Then for all
$B\in \gS_{q_1}(\gH)$ we have $B\in\gS_{q_2}(\gH)$ and
  \[\begin{split}
    \tr\!(B|D_0|^{1/2}\gamma^{(p_1)}_\infty|D_0|^{1/2})
    \!=\!\lim_{n\to\infty}
    \tr\!(B|D_0|^{1/2}\gamma_n|D_0|^{1/2})
    \!=\!\tr\!(B|D_0|^{1/2}\gamma^{(p_2)}_\infty|D_0|^{1/2})\;.
  \end{split}\]
Thus, for all $B\in \gS_{q_1}(\gH)$, we have
 $$
  \tr(B |D_0|^{1/2}(\gamma^{(p_1)}_\infty
  -\gamma^{(p_2)}_\infty)|D_0|^{1/2})=0\;.
 $$
This yields $\gamma^{(p_1)}_\infty=\gamma^{(p_2)}_\infty$.

Therefore, in the other parts of this proof, we will always assume
we have chosen a subsequence of $\gamma_n$ such that, for the
considered $p$, the weak-limits in $(F, \|.\|_{F, p})$ exist and
coincide.

\textbf{Step 2. Lower Semi-Continuity:} We will now prove that
taking the limit decreases the energy, i.e.
$\liminf_{n\to\infty}\cE_m(\gamma_n)\geq\cE_m(\gamma_\infty)$. We
define $\Lambda_m:= \chi_{[0,m]}(\Dd)$ and
$\Lambda_m':=\Ld_+-\Lambda_m$ and split the energy functional. We
treat the various terms separately
\[\begin{split}
    \cE_{m}(\gamma_n) = & T_1(\gamma_n) + T_2(\gamma_n)
    + T_3(\gamma_n) + T_4(\gamma_n) + T_5(\gamma_n) \\
    := & \tr(\Lambda_m'(\Dd-m )\Lambda_m'\gamma_n) +
    \tr(\Lambda_m(\Dd-m)\Lambda_m\gamma_n)- \alpha\tr(\fid\gamma_n)\\
    &+ \alpha\tr(\Xd \gamma_n) + \alpha Q(\gamma_n,\gamma_n)\;.
\end{split}\]


  \textbf{ Step 2.1.} Fix a basis $(e_\ell)_{\ell \in \mathbb N}$ of $\gH_+$, each element being in $H^1(G)$. Then
  \begin{eqnarray*}
   T_1(\gamma_n) & := & \tr \left ( \Lambda_m' (\Dd-m
   )\Lambda_m'\gamma_n\right) \\ & = &\tr \left ( (\Lambda_m' (\Dd-m
   )\Lambda_m')^{1/2}\gamma_n(\Lambda_m' (\Dd-m
   )\Lambda_m')^{1/2}\right) \\ & = &\sum_k(e_k,(\Lambda_m' (\Dd-m
   )\Lambda_m')^{1/2}\gamma_n \underbrace{(\Lambda_m' (\Dd-m
   )\Lambda_m')^{1/2}e_k}_{f_k:=})\\
   &=&\sum_k\tr\big(\underbrace{\big||D_0|^{-1/2}f_k\big\rangle\big\langle
   |D_0|^{-1/2}
   f_k\big|}_{H_k:=}|D_0|^{1/2}\gamma_n|D_0|^{1/2}\big)\;.
\end{eqnarray*}
Obviously, $H_k$ is a non-negative Hilbert-Schmidt operator.  Thus, applying
first Fatou's lemma and then using \eqref{eq:weak},  we get
\begin{equation}
\begin{array}{ll}
\liminf_{n\to\infty} \tr \,\left ( \Lambda_m'
(\Dd-m )\Lambda_m'\gamma _n
  \right ) & =
\liminf_{n\to\infty}\sum_k \tr(H_k |D_0|^{1/2}\gamma_n |D_0|^{1/2})\\
& \geq
  \sum_k\liminf_{n\to\infty}\tr(H_k|D_0|^{1/2}\gamma_n|D_0|^{1/2})\\
  & =  \sum_k\tr(H_k|D_0|^{1/2}\gamma_\infty|D_0|^{1/2})\\
   & = \tr \left ( \Lambda_m' (\Dd-m )\Lambda_m'\gamma _{\infty}
   \right)\;,
\end{array}
\end{equation}
which proves $\liminf_{n\to\infty}T_1(\gamma_n)\geq T_1(\gamma_\infty)$.


\textbf{ Step 2.2.} Because $T_2$ is continuous in the $\|\cdot\|_{F,2}$-norm
(Lemma \ref{sec:some-spectr-prop-1}) the claim follows for $T_2$.


\textbf{Step 2.3.} Since $\fid\in L^4(\rz^3)$ (Lemma
\ref{l3tierce}), we have, by \cite[Theorem~4.1]{Simon1979T}, that
$|D_0|^{-1/2} \fid |D_0|^{-1/2}\in \gS_4(\gH)$. By H\"older
inequality, this implies that $T_3$ is continuous in the
$\|\cdot\|_{F,4/3}$-norm, and
$\lim_{n\rightarrow+\infty}T_3(\gamma_n)=T_3(\gamma)$.


\textbf{ Step 2.4.} We would like to prove
\begin{equation}\label{eq:sc4}
 \liminf_{n\rightarrow\infty} Q[\gamma_n,\gamma_n]
 \geq Q[\gamma_\infty,\gamma_\infty] \;.
\end{equation}
To that end, we will first show
\begin{equation}\label{poinwise-prop.1}
\lim_{n\rightarrow\infty}\gamma_n(x,y) = \gamma_\infty(x,y)\;,
\end{equation}
for a.e. $(x,y)\in G^2$, and
\begin{equation}\label{poinwise-prop.2}
\lim_{n\rightarrow\infty}\gamma_n(x,x) = \gamma_\infty(x,x)\;,
\end{equation}
for a.e. $x\in G$.

Now, $(\gamma_n)$ is a bounded sequence in $\gS_2(\gH)$. Again, we
can extract a subsequence such that $\gamma_n$ converges weakly to
$\tilde\gamma_\infty$ in $\gS_2(\gH)$. Using \eqref{eq:weak}, we
get for all $B\in\gS_2(\gH)$
 \begin{eqnarray*}
  \tr(B\tilde\gamma_\infty)& =
  &\lim_{n\rightarrow\infty}\tr(B\gamma_n) = \lim_{n\rightarrow\infty}
  \tr\big(\underbrace{|D_0|^{-\frac12}B|D_0|^{-\frac12}}_{\in\gS_2(\gH)}
  |D_0|^{\frac12}\gamma_n|D_0|^{\frac12} \big)\\ & = & \tr
  \big(|D_0|^{-\frac12}B|D_0|^{-\frac12}
  |D_0|^{\frac12}\gamma_\infty|D_0|^{\frac12}\big) =
  \tr(B\gamma_\infty)\;.
 \end{eqnarray*}
Thus $\tilde\gamma_\infty = \gamma_\infty$. In particular we have
\begin{equation}
\label{weak}
  \gamma _n (\cdot ,\cdot )\to \gamma _{\infty}(\cdot ,\cdot )\;,
\end{equation}
weakly in $L^2(G\times G)$.

Using the spectral decomposition of the $\gamma_n$'s, we may write
each $\gamma _n(x,y)$ as
 $$
  \gamma _n(x,y)=\sum_{i=1}^q
  u_i^{(n)}(x)\overline{u_i^{(n)}(y)}\;,
 $$
where each sequence $(u_i^{(n)})_{n\in\nz}$ ($i=1,\ldots,q$) is an
orthonormal family in $\in H^{1/2}(G) \cap\Ld_+\gH$.  Since the
sequence $(\tr (|p|\gamma_n))_{n\in\nz}$ is bounded, it follows
that, for each $i\in\{1,\,\ldots, {q}\}$, the sequence
$(u_i^{(n)})_{n\in\nz}$ is bounded in $H^{1/2}(G)$.

Therefore, applying \cite[Theorem~16.1]{LionsMagenes1972}, for all
$\chi\in C_0^\infty(\rz^3)$ and $i\in\{1,\ldots,q\}$, there exists
a subsequence of $(u_i^{(n)})_{n\in\nz}$ -- also denoted
$(u_i^{(n)})_{n\in\nz}$ -- such that $(\chi u_i^{(n)})$ converges
strongly in $L^2(G)$.

Thus, after
extraction of a subsequence of $\left(u_1^{(n)},u_2^{(n)},
\ldots,u_q^{(n)}\right)_{n\in\nz}$, denoted again by
$\left(u_1^{(n)},u_2^{(n)}, \ldots,u_q^{(n)}\right)_{n\in\nz}$, we
obtain, for all $i=1,\ldots,q$ and for almost every $x\in G$,
\begin{equation}\label{pointwise-conv}
u_i^{(n)}(x)\to u_i^{(\infty )}(x)\;.
\end{equation}
Consequently, we obtain
\begin{equation}
\label{eq:b}
  \gamma _n(x , y )\to \beta (x ,y) := \sum_{i=1}^q u_i^{(\infty )}(x)\,
\overline{u_i^{(\infty )}(y)}\;,
\end{equation}
almost everywhere in $G^2$. Now from \eqref{weak} and \eqref{eq:b}
it follows by standard arguments that $\gamma _{\infty}(x,y) =
\beta (x,y)$ almost everywhere in $G\times G$. Thus
$\gamma_n(x,y)$ converges almost everywhere to
$\gamma_\infty(x,y)$. This together with \eqref{eq:b} implies
\eqref{poinwise-prop.1}.

The above also immediately implies that
\begin{equation}\label{eq1}
\gamma_\infty = \sum_{i=1}^q |u_i^{(\infty)} \rangle\langle
u_i^{(\infty)}|\;,
\end{equation}
proving \eqref{poinwise-prop.2}.

Applying Fatou's lemma to the pointwise positive functions\break
$(x,y)\mapsto \gamma_n(x,x)\gamma_n(y,y) - |\gamma_n(x,y)|^2$ and using in addition
\eqref{poinwise-prop.1} and \eqref{poinwise-prop.2} yields
\begin{eqnarray*}
        \liminf\limits_{n\to\infty} Q[\gamma_n,\gamma_n]
&\geq &
        \int\liminf\limits_{n\to\infty}\frac{\gamma_n(x,x)
        \gamma_n(y,y)-|\gamma_n(x,y)|^2}
         {| \bx - \by|} \rd x\, \rd y \\
& = &
         \int \frac{\gamma_{\infty}(x,x)\gamma_{\infty}(y,y)-
         |\gamma_{\infty}(x,y)|^2}{| \bx - \by|} \rd x\, \rd y
=
      Q[ \gamma_\infty , \gamma_\infty]\;,
\end{eqnarray*}
which proves \eqref{eq:sc4}.


\textbf{ Step 2.5.} Since $\fid\in L^4(\rz^3)$ (Lemma
\ref{l3tierce}) and $\Xd\leq\fid$ (by Lemma~\ref{lem:Xsmallphi} and the
positivity of $\Xd$ and $\delta$),  we have
$|\Dd|^{-1/2}\Xd|\Dd|^{-1/2}\in\gS_4$. Thus $T_5$ is continuous in
the $\|\cdot\|_{F,4/3}$-norm (by H\"older inequality). Therefore
$$\lim_{n\rightarrow+\infty} T_5(\gamma_n) = T_5(\gamma_\infty)\;.$$
This concludes the proof of Theorem~\ref{existence}.
\end{proof}
\begin{theorem}\label{sec:existence-cor} Assume $0\leq\delta\in F$,
$q\in \nz$, $q\leq Z$ and
  $$
   \pi\alpha(1/4
   +\max\{\tr\delta,q\})<(d-4\alpha\tr\delta) .
   $$
   Then the functional $\cE|_{\Sd_{\partial q}}$ (see the
   Definition \eqref{def1} and \eqref{set:1pdmdq}) has a minimizer.
\begin{proof}
Using lemma~\ref{lmodulus} and the assumptions, we obtain
 \begin{equation}\label{sec:existence-ineq-1}
   \pi \alpha (\frac14 +\max\{\tr\delta,q\}) < \cda .
 \end{equation}
Therefore, by Theorem~\ref{sec:minimization-energy-2}, the functional
$\cE_m$ is bounded below on $\Sd_q$.

Lemmata~\ref{diff-proj}, \ref{sec:reduct-unren-dens} and
\ref{sec:reduct-proj-1} together with Remark~\ref{rem:charge}
imply that it suffices to minimize over positive $\gamma$'s in
$\Sd_q$ that can be written as a sum of a projection and a rank
one operator. Moreover, Inequality \eqref{sec:existence-ineq-1} and
Lemma~\ref{gs} permit to apply  Lemma~\ref{t1}, which, together with the above and the fact that $q$ is
an integer yields
 \begin{equation}\label{sec-6-eq-1}
   \inf \{\cE_m(\gamma) | \gamma\in \Sd_q \} =
   \inf \{\cE_m(\gamma) | \gamma\in \Sd_{\partial q}, \gamma^2 =\gamma\}.
 \end{equation}
Thus, applying Theorem~\ref{existence} and again Lemma~\ref{t1}
shows that $\cE_m|_{\Sd_q}$ has a minimizer in $\Sd_{\partial q}$.
Therefore the functional $\cE|_{\Sd_{\partial q}}$ also has a
minimizer.
\end{proof}

\end{theorem}

\section{Properties of the  Minimizers: No-Pair Dirac-Fock
  Equations\label{s7}}

\subsection{Infinitesimal Perturbations of Projections}
\label{sec:infin-pert-proj}
In this section, we will prove that all minimizers $\gamma$ fulfill a no-pair
Dirac-Fock equation. We first need to state preliminary results. The first was
already used in the adiabatic theory (see Nenciu \cite{Nenciu2002} and
references therein).
\begin{lemma}[Nenciu] \label{l4}
  Given an orthogonal projection $P_0$, its orthogonal complement $P_0^\perp =
  1-P_0$, a bounded operator $A$, and $\epsilon\in\rz$, with $4 |\epsilon|\,
  \|A\| < 1 $, there exists an operator $B_\epsilon$ with $\| B_\epsilon\|
  \leq 4 \, \|A\|^2$ such that
\begin{equation}\label{lem:l4-def-beps}
    P_\epsilon = P_0 + \epsilon \left(P_0 A P_0^\perp +
    P_0^\perp A^\ast P_0\right) + \epsilon^2 B_\epsilon
\end{equation}
is an orthogonal projection.
\end{lemma}

\begin{proof}
  We set
  \begin{equation}
    \label{eq:nenciu1}
    P_\epsilon:= \frac1{2\pi i} \oint_{|z-1|=\frac12}
    \frac1{z-P_0-\epsilon a}\rd z ,
  \end{equation}
  where
  $$
  a:=P_0 A P_0^\perp + P_0^\perp A^*P_0\; .
  $$
  We observe that $\|a\| \leq \|A\|$. Therefore under the assumption
  $4\epsilon\|A\|<1$, we obtain that $\sigma(P_0 + \epsilon a)\subset (-1/4,
  1/4)\cup (3/4, 5/4)$. Thus, by the holomorphic functional calculus,
  $P_\epsilon$ is the projector onto the eigenspace of $(P_0+\epsilon a)$
  corresponding to $(3/4, 5/4)$.
\begin{equation}\label{eq:nenciu2}
 \begin{split}
   P_\epsilon = &\, P_0 + \frac\epsilon{2\pi i} \oint_{|z-1|=\frac12}
   \frac1{z-P_0} a \frac1{z-P_0}\, \rd z\\
   & - \frac{\epsilon^2}{2\pi i}\oint_{|z-1|=\frac12} \frac1{z-P_0}a
   \frac1{z-P_0-\epsilon a}a \frac1{z-P_0}\,\rd z \;.
\end{split}
\end{equation}
Since $P_0$ is an orthogonal projection, there exists a basis
$(e_j)_{j\in\nz}$ of $\gH$ and $I\subset\nz$ such that
$$
P_0 = \sum_{j\in I} |e_j\rangle \langle e_j| .
$$
Note also for later purpose that
\begin{equation}\label{eq:resolvent2}
 \frac{1}{z-P_0} = \frac{1}{z-1}\sum_{j\in I}  |e_j\rangle\langle e_j|
 + \frac{1}{z}\sum_{j\in\nz\setminus I}  |e_j\rangle \langle e_j| =
 \frac{1}{z-1}P_0 + \frac{1}{z}P_0^\perp \;.
\end{equation}
Using Cauchy's residue Theorem and \eqref{eq:resolvent2}, we obtain
\begin{equation}\label{eq:nenciu3}
\begin{split}
  \frac{1}{2 i\pi}{\oint_{|z-1| = \frac12} \frac{1}{z-P_0} a \frac{1}{z-P_0}
    \rd z} = & \frac{1}{2 i\pi}\oint_{|z-1| = \frac12} \left(P_0^\perp a P_0+
    P_0 a
    P_0^\perp\right) \frac{\rd z}{z(z-1)}\\
  = &\ P_0^\perp a P_0 + P_0 a P_0^\perp = a \;.
\end{split}
\end{equation}
This proves that the second summand in the right hand side of
\eqref{eq:nenciu2} is equal to $\epsilon a$. This leads us to introduce~:
\begin{equation}\label{defBeps}
B_\varepsilon:=  - \frac{1}{2\pi i}\oint_{|z-1|=\frac12} \frac1{z-P_0}a
\frac1{z-P_0-\epsilon a}a \frac1{z-P_0}\,\rd z\;.
\end{equation}
Since $\| a \| \leq \| A \|$ and $\sigma(P_0 + \epsilon a)\subset (-1/4,
1/4)\cup (3/4, 5/4)$, we have
\begin{eqnarray}
 \lefteqn{\left\|\oint_{|z-1|=\frac12} \frac1{z-P_0}a
 \frac1{z-P_0-\epsilon a}a \frac1{z-P_0}\,\rd z\right\| } & &
 \nonumber\\
 & \leq & \pi \|a\|^2 \sup_{|z-1|=\frac12} \| \frac{1}{z-P_0}\|^2
 \sup_{|z-1|=\frac12} \| \frac{1}{z-P_0-\epsilon a}\|
 \leq 16 \pi \|A\|^2 \label{estut}.
\end{eqnarray}
This, together with \eqref{eq:nenciu2} and \eqref{eq:nenciu3}, proves
Lemma~\ref{l4}.
\end{proof}

In the case when $\gamma$ is an orthogonal projection with range in $\gH_+$,
we apply Lemma \ref{l4} with $P_0 = \gamma$, $P_\epsilon = \gamma_\epsilon$
(given by \eqref{lem:l4-def-beps}).
\begin{lemma}\label{proj-stab} Assume $0\leq\delta\in F$ and
  $\gamma$ an orthogonal projection in $\Sd_{\partial q}$, $q\in \nz$. Then,
  for operators $A$ such that $\Ld_+ A\Ld_+ =A$, $|D_0|A\in\gS_1(\gH)$ and
  $\epsilon$ sufficiently close to zero, $\gamma_\epsilon$ is again an
  orthogonal projection in $\Sd_{\partial q}$.
\end{lemma}
\begin{proof}
  By construction, since $\Ld_+ A\Ld_+ =A$, we have
  $$
  {\Ld_-\gamma_\epsilon\Ld_-}=\Ld_+
  \gamma_{\epsilon}\Ld_-=\Ld_-\gamma_{\epsilon}\Ld_+=0\;.
  $$
  Moreover, $\gamma_\epsilon^2 = \gamma_\epsilon$, thus
  $-\Ld_-\leq\gamma_\epsilon\leq \Ld_+$ and $\Ld_-\gamma_\epsilon\Ld_+=0$.

  The trace condition is obviously fulfilled since $\tr\gamma_\epsilon$
  depends continuously on the parameter $\epsilon$. That $D_0\gamma_\epsilon$
  is trace class follows immediately from the explicit expressions for the
  difference $\gamma_\epsilon-\gamma$ in \eqref{eq:nenciu2}, from
  \eqref{eq:resolvent2} and the assumptions on $A$.
\end{proof}

\subsection{Minimizers are Projections}
\label{sec:minim-are-proj}

\begin{theorem}\label{sec:min-are-proj-1}
 Assume that $0\leq\delta\in F$, $q\in\nz$, and
 $$0<\pi\alpha(1/4+\max\{\tr\delta,q\})< \cda\;.$$
 If $\gamma$ is a
 minimizer of $\cE|_{\Sd_{\partial q}}$, then
 $\gamma=\gamma^*=\gamma^2=\Ld_+\gamma\Ld_+$.
\end{theorem}
\begin{proof}
The proof of $\Ld_+\gamma\Ld_+=\gamma$ is a consequence of
Lemma~\ref{diff-proj}. The proof of $\gamma^2=\gamma$ follows
exactly the lines of Lemma~\ref{sec:reduct-proj-1} except that the
iteration of the process is superfluous here.
\end{proof}

\subsection{Minimizers Fulfill the No-Pair Dirac-Fock Equations}
\label{sec:minim-fulf-no}

Eventually we derive the Euler equations for the minimizer of the
energy.
\begin{theorem} \label{th:hf}
Assume $q\in\nz_0$ and $\gamma$ is an orthogonal projection
minimizing $\cE$ in $\Sd_{\partial q}$. Then $\gamma$ commutes
with the no-pair Dirac-Fock operator $\Ld_+D^{(\gamma)}\Ld_+$,
i.e.,
\begin{equation}\label{eq:hf}
    \left[\gamma, \Ld_+D^{(\gamma)}\Ld_+\right] =0.
\end{equation}
\end{theorem}
\begin{proof}
Let $A\in\mathcal{B}(\gH)$ such that
\begin{equation}\label{Astar}
     \Dd A^* \in \mathcal{B}(\gH)\;.
\end{equation}
Then, for $\epsilon$ sufficiently close to zero, the projector
\begin{equation}
    \label{eq:euler1}
    \gamma_\epsilon := \gamma
    + \epsilon a + \epsilon^2 B_\epsilon \;,
\end{equation}
with
\begin{equation}
\label{eq:deriv-3} a=\gamma \Ld_+A \Ld_+ \gamma^\perp + \gamma^\perp \Ld_+A^*
\Ld_+ \gamma \;,
\end{equation}
and $B_\epsilon$ given by \eqref{defBeps} (with $P_0$ replaced by
$\gamma$), belongs to $\Sd_{\partial q}$. Moreover
\begin{multline}\label{eq:derivative}
  \cE(\gamma_\epsilon) - \cE(\gamma)\\
  = \epsilon\{ \tr(D_Z a) + 2\alpha \Re Q[\gamma, a]\} +
  \epsilon^2 \{
  \tr(D_Z B_\epsilon) + 2\alpha\Re Q[\gamma, B_\epsilon] +\alpha
  Q[a+\epsilon B_\epsilon , a+\epsilon B_\epsilon]\} .
\end{multline}
We want to show that the last term is $o(\epsilon)$. By
\eqref{eq:q1}, \eqref{eq:3} and Lemmata~\ref{l3tierce} and
\ref{lem:Xsmallphi},  it is sufficient to show that there exists a
constant $\const <\infty$ such that, for all $\epsilon\in (-1, 1)$,
\begin{equation}\label{eq:max1}
\max\{ \|B_\epsilon \|_1 , \|a\|_1, \|\Dd B_\epsilon\|_1, \|\Dd
a\|_1 \} <c\;.
\end{equation}
We first have
 \begin{equation}\label{eq:max1-1}
 \begin{split}
   \| \Dd a \|_1 & \leq \|\Dd \gamma\Ld_+ A \Ld_+
   \gamma^\perp \|_1 +
   \| \Dd \gamma^\perp \Ld_+ A^* \Ld_+ \gamma\|_1 \\
   & \leq \| \Dd \gamma\|_1\, \|A\| +
   \| \Dd\gamma\|\, \|A^*\|\, \|\gamma\|_1 + \| \Dd A^* \|\,
   \|\gamma\|_1 <\const \;,
 \end{split}
 \end{equation}
since $\gamma\in F$ and \eqref{Astar} is assumed. We also have
\begin{equation}\nonumber
\begin{split}
  \| \Dd B_\epsilon\|_1&= \frac1{2\pi}\|\oint_{|z-1|=\frac12} \Dd
  \frac1{z-\gamma}a \frac1{z-\gamma-\epsilon a}a \frac1{z-\gamma}
  \rd z\|_1\\
  &\leq \frac1{2\pi}\!\int_0^{2\pi}\!\!\!\!\| \Dd
  \frac1{1+\frac12e^{i\varphi}-\gamma}a \|_1\,
  \|\frac1{1+\frac12e^{i\varphi}-\gamma-\epsilon a}a
  \frac1{1+\frac12e^{i\varphi}-\gamma}\|\,\rd \varphi \\
  &\leq \const_1 \,||a||\,\int_0^{2\pi}\| \Dd
  \left(\frac1{1+\frac12e^{i\varphi}-1}\gamma +
    \frac1{1+\frac12e^{i\varphi}}(1-\gamma)\right)
  a \|_1 \rd \varphi \\
  & \leq \const_2 \,||a|| \, \left( \|\Dd \gamma\|_1 \|a\|
  + \|\Dd a\|_1 \right)\;,
\end{split}
\end{equation}
where the constant $\const_2$ is uniform in $\epsilon$ for
$\epsilon$ close to zero. We also have used  \eqref{eq:resolvent2}.
Using \eqref{eq:max1-1}, and $\gamma\in F$ yields
\begin{equation}\label{eq:max1-2}
 \| \Dd B_\epsilon\| < \const\;.
\end{equation}
Similarly to the above, we show
 \begin{equation}\label{eq:max1-3}
   \| a \|_1 < \const \quad\mbox{and}\quad \|B_\epsilon\|_1<\const\;.
 \end{equation}
Inequalities \eqref{eq:max1-1}, \eqref{eq:max1-2} and
\eqref{eq:max1-3} yield \eqref{eq:max1}.

Since $\mathcal{E}(\gamma_\epsilon)-\mathcal{E}(\gamma) \geq 0$,
whatever the sign of $\epsilon$ is, we conclude that the term
linear in $\epsilon$ in \eqref{eq:derivative} has to vanish
 \begin{equation}\label{eq:deriv-2}
   \tr(D_Z a) + 2\alpha\Re Q[\gamma, a] = \tr(D^{(\gamma)} a) = 0\; .
 \end{equation}
Thus, for all $A$ satisfying \eqref{Astar}, equalities \eqref{eq:deriv-3} and
\eqref{eq:deriv-2} and the fact that $[\Ld_+, \gamma]=0$ (since
$\gamma$ is an orthonormal projection in $\Sd_{\partial q}$) yield
 $$
    \tr( \gamma^\perp \Ld_+D^{(\gamma)} \Ld_+ \gamma A )
    + \tr( \gamma \Ld_+ D^{(\gamma)} \Ld_+ \gamma^\perp A^*) = 0\;.
 $$
Replacing $A$ by $i A$, we first obtain~:
$$ \tr( \gamma^\perp \Ld_+D^{(\gamma)} \Ld_+ \gamma A )=
     \tr( \gamma \Ld_+ D^{(\gamma)} \Ld_+ \gamma^\perp A^*) = 0\;.
$$
Since $A$ can be taken in  the set of rank one operators of the
form $| u\rangle\,\langle v|$, with $u$ and $v$ in
$C_0^\infty(\rz^3)\otimes\cz^4$, we obtain
 $$
  \gamma^\perp \Ld_+ D^{(\gamma)} \Ld_+ \gamma =
  \gamma \Ld_+ D^{(\gamma)} \Ld_+ \gamma^\perp = 0\;,
 $$
which yields \eqref{eq:hf}.
\end{proof}
This result can be also written as follows
\begin{theorem}\label{th:proj}
  Assume that $\gamma$ is an orthogonal projection minimizing $\cE$ in
  $\Sd_{\partial q}$. Then there exist $q$ normalized spinors
  $\psi_1,...,\psi_q\in \Ld_+(\gH)\cap \mathfrak{D}(D_Z)$ such that
 \begin{equation}\label{th7:proj-form}
  \gamma = \sum_{i=1}^q |\psi_i\rangle\,\langle\psi_i|\;,
 \end{equation}
and
 \begin{equation}\label{th7:DF-ev}
 \Ld_+ D^{(\gamma)} \Ld_+ \psi_i =
 \epsilon_i\psi_i,\ \mbox{ for } i=1,\,\ldots,\, q\;,
 \end{equation}
with $\epsilon_1,...,\epsilon_q\in [0,1]$.
\end{theorem}
\begin{proof}
  The proof is immediate since the range of $\gamma$ is finite
  dimensional reducing it to the simultaneous diagonalization of two
  commuting Hermitian matrices.
\end{proof}
One may characterize the eigenvalues $\epsilon_1,\dots , \epsilon_q$
more precisely following Bach et al  \cite{Bachetal1994}:
\begin{theorem}\label{unfilled-shell}
[There are no unfilled shells in no-pair Dirac-Fock theory]
Under the same assumptions of Theorem \ref{th:proj},
  $\epsilon_1,...,\epsilon_q$ are the $q$ lowest eigenvalues of
  $\Ld_+\Dd\Ld_+$. Moreover, if $\epsilon_{q+1}$ denotes the
  ($q+1$)-th eigenvalue (counting multiplicities) of the no-pair
  Dirac-Fock operator $\Ld_+D^{(\gamma)}\Ld_+$, we have, for all
  $i=1,\cdots,q$, the strict inequality $\epsilon_i < \epsilon_{q+1}$.
\end{theorem}
\begin{proof}
  Assume by contradiction  that there exists a normalized eigenspinor
  $v$ of $\Ld_+D^{(\gamma)}\Ld_+$ with eigenvalue $\epsilon$ such that
  $\epsilon \leq \max\{\epsilon_1,...,\epsilon_q\}$ and not in the
  range of $\gamma$. Then, for a normalized eigenvector $u$ of
  $\gamma$ with $\langle u,\Ld_+ D^{(\gamma)}\Ld_+ u\rangle \geq
  \epsilon$ and for $\gamma':=\gamma -|u \rangle\,\langle u|\, +\,
  |v\rangle\,\langle v|$, we have
\begin{eqnarray*}
 \lefteqn{\cE^{(\delta)}(\gamma') -
 \cE^{(\delta)}(\gamma)}  & & \\
 & = & \langle v, D_Z v\rangle - \langle u , D_Z u \rangle
 + 2\alpha\Re Q\left[\gamma, |v\rangle\,\langle v| - |u \rangle\,
 \langle u |\right] \\
 & & + Q\left[|v\rangle\,\langle v| - |u \rangle\,
 \langle u |, \, |v\rangle\,\langle v| - |u \rangle\,
 \langle u|\right] \\
 & < & \langle v, D^{(\gamma)} v\rangle -
 \langle u, D^{(\gamma)} u\rangle \leq 0 ,
\end{eqnarray*}
where -- as in the proof of Lemma \ref{sec:reduct-proj-1} -- we used
 \eqref{QSS}, with $S = |v\rangle\,\langle v| - |u\rangle\,
  \langle u|$. This gives a contradiction to the property that $\gamma$
 is the minimizer.
\end{proof}


\section{Outlook}

In this final section, we first express the energy as a functional of the
unrenormalized density matrix $\Gamma$ and the Dirac sea $\Lambda_-$. This has
the advantage that the dependence of the energy on $\Lambda_-$ becomes
explicit and that the constraining condition $-\Lambda_- \leq \gamma \leq
\Lambda_+$ becomes $0\leq \Gamma\leq 1$, i.e.  independent of $\Lambda_-$.
Throughout this section we will use the notation $\Lambda:=\Lambda_-$.

For a given $q\in\nz$, let us define the set
\begin{eqnarray*}
  \Upsilon_q & := & \Big\{ (\Gamma, \Lambda)\in \gB(\gH)^2\ |\
  \Gamma, \Lambda\ \mbox{orth. proj.},
  (\Gamma-\Lambda)\in\gS_1, \,
  \tr\left(\Gamma - \Lambda\right) = q, \\
  & &
  \ \ \ D_Z(\Gamma - \Lambda)\in\gS_1,
  [D_Z, \Lambda]\in\mathcal{B}(\gH), [D_Z, \Gamma]\in\mathcal{B}(\gH) \Big\}
\end{eqnarray*}
and the following functional on $\Upsilon_q$:
\begin{equation*}
 \gE(\Gamma, \Lambda) :=
 \tr\left((D_Z - m) (\Gamma-\Lambda)\right)
 + \alpha Q[\Gamma - \Lambda, \Gamma- \Lambda] \;,
\end{equation*}
where $Q[\cdot , \cdot ]$ is defined in \eqref{def-Q}. Note that if $\Lambda =
\Ld_-$ for some $\delta\in F$, and if $\left(\Gamma-\Lambda\right) \in \Sd$,
then we have $\gE( \Gamma, \Lambda ) = \cE (\Gamma- \Lambda)$.

\begin{theorem}\label{wbr}
  Assume that $(\Gamma, \Lambda)\in \Upsilon_q$ is a critical point of $\gE$.
  Then, with $\gamma:=\Gamma-\Lambda$,
\begin{equation}\label{eq:wbr}
  [D^{(\gamma)}, \Gamma] = [D^{(\gamma)}, \Lambda] = 0 .
\end{equation}
\end{theorem}
\begin{proof} For all $\epsilon\in\rz$ and
  $A\in\gS_1(\gH)$ such that~:
\begin{equation}\label{eq:condition5-1}
 D_Z A \in \gS_1(\gH)\,,\, D_Z A^* \in
  \gS_1(\gH)\,,\,4|\epsilon| \|A\| <1\,, \,8q\epsilon^2\|A\|^2 <1\mbox{ and
  }\epsilon^2 \|A\|_1<1\;,
\end{equation}
we define (see Lemma \ref{l4}) the orthogonal projector~:
\begin{equation}\label{eq:perturbation}
 \Lambda_\epsilon = \frac{1}{2\pi i}\oint_{|z-1| = \frac12}
 \frac{1}{z - \Lambda - \epsilon a}\; \rd z\;,
\end{equation}
where $a = \Lambda A \Lambda^\perp + \Lambda^\perp A^* \Lambda$. From
Lemma~\ref{l4}, we get the decomposition
\begin{equation}\label{eq:perturbation2}
 \Lambda_\epsilon = \Lambda +
 \epsilon a + \epsilon^2 B_\epsilon \;,
\end{equation}
with $P_0$ replaced by $\Lambda$. \\ Let us prove that $(\Gamma,
\Lambda_\epsilon)$ belongs to $\Upsilon_q$.

We first show $\Gamma-\Lambda_\epsilon\in\gS_1(\gH)$. We have
\begin{equation}\label{eq:app1}
  \| a \|_1 \leq 2 \|A\|_1 <\infty \;,
\end{equation}
and, as in the proof of \eqref{estut}, we get
\begin{equation}\label{eq:app2}
   \|B_\epsilon \|_1 \leq 16 \pi \|A\|\,\|A\|_1 \;.
\end{equation}
Therefore, $\Gamma-\Lambda_\epsilon = \Gamma-\Lambda + \epsilon a +
\epsilon^2B_\epsilon\in\gS_1(\gH)$.

We next establish $D_Z\Lambda_\epsilon\in\gS_1(\gH)$. Since $(\Gamma,
\Lambda)\in \Upsilon_q$, $[D_Z, \Lambda]$ is a bounded operator and $A$, $D_Z
A\in\gS_1(\gH)$. Thus
\begin{equation}
\begin{split}\label{app-est1}
  \| D_Z \Lambda A \Lambda^\perp \|_1 & \leq \|\Lambda D_Z A \Lambda^\perp\|_1
  + \|\, [\,D_Z\,, \,\Lambda\,] A \Lambda^\perp
  \|_1\\
  & \leq \|D_Z A \|_1 + \| \,[\,D_Z\,,\, \Lambda\,]\,\|~\|A\|_1 <\infty \;.
\end{split}
\end{equation}
Similarly, we can prove
\begin{equation}\label{app-est2}
  \|D_Z \Lambda^\perp A^* \Lambda \|_1\
  \leq  \|D_Z A^* \|_1 +  \|\, [\,D_Z\,,\, \Lambda^\perp\,]\,\|~\|A^*\|_1
  <\infty  \;,
\end{equation}
which implies, together with \eqref{app-est1} that there exists a constant $c$
such that
\begin{equation}\label{app-est2.5}
 \| D_Z a \|_1 < c \;.
\end{equation}
Using again that $[\,D_Z\,,\, \Lambda\,]$ is bounded and
Formula~\eqref{eq:resolvent2}, valid with $\Lambda$ instead of $P_0$ and
$\Lambda^\perp$ instead of $P_0^\perp$, we get the existence of a constant $c$
such that for all $\epsilon$ small enough
\begin{equation}\label{app-est3}
\| D_Z  B_\epsilon \|_1 < c\; .
\end{equation}
Inequalities \eqref{app-est2.5} and \eqref{app-est3} yield $D_Z
\Lambda_\epsilon\in\gS_1(\gH)$.

Now, from \eqref{app-est2.5}, \eqref{app-est3} and $[D_Z, \Lambda_\epsilon] =
[D_Z, \Lambda] + [D_Z, \epsilon a+ \epsilon^2 B_\epsilon]$, we obtain
$[\,D_Z\,,\, \Lambda_\epsilon\,]\in\mathcal{B}(\gH)$.

We finally prove that $\tr\left(\Gamma-\Lambda_\epsilon\right) =q$. For that
purpose, we first note that, due to Effros \cite{Effros1989} (see also Avron
et al \cite[Theorem~4.1]{Avronetal1994}), and since from the above
$\Gamma-\Lambda_\epsilon\in\gS_1(\gH)$, and both $\Gamma$ and
$\Lambda_\epsilon$ are projections, we have $\tr\left(\Gamma-
  \Lambda_\epsilon\right) \in\gz$. Furthermore, from \eqref{eq:perturbation2},
\eqref{eq:app1} and \eqref{eq:app2} we get
$$
\|(\Gamma - \Lambda) - (\Gamma-\Lambda_\epsilon)\|_1 =\mathcal{O}(\epsilon)
\;.
$$
Since $\tr(\Gamma-\Lambda_\epsilon)$ is an integer and
$\tr(\Gamma-\Lambda)=q$, this yields, for $\epsilon$ small enough,
$\tr\left(\Gamma- \Lambda_\epsilon\right) = q$. This concludes the proof that
$(\Gamma,\Lambda_\epsilon)\in \Upsilon_q$.

Let us now prove that for $\gamma:=\Gamma-\Lambda$ we have
\begin{equation}\label{eq:app-comm}
  [D^{(\gamma)}, \Lambda] = 0 \;.
\end{equation}
Since $(\Gamma, \Lambda)$ is a critical point of $\gE$, for all
$A\in\gS_1(\gH)$ satisfying \eqref{eq:condition5-1}, we have
\begin{equation}\label{eq:app-crit1}
 \frac {\partial\gE( \Gamma,\Lambda_\epsilon)}
 {\partial\epsilon}\Big|_{\epsilon =0} = 0 \; ,
\end{equation}
where $\Lambda_\epsilon$ is defined by \eqref{eq:perturbation2}.  On the other
hand, we have
\begin{equation}\label{eq:bound-5}
 \begin{array}{ll}
   \gE( \Gamma,\Lambda_\epsilon) & \!\!\!\! =
   \tr\left((D_Z-m)\gamma\right)
   \!+\! \alpha Q[\gamma, \gamma]
   - \epsilon \Big\{ \tr\left((D_Z-m) a  \right)
    + 2\alpha\Re Q[a, \gamma]\Big\}\\
   &\ +\, \epsilon^2 \Big\{ \tr\left((D_Z-m)B_\epsilon\right)+
   \alpha Q[a + \epsilon B_\epsilon, a + \epsilon B_\epsilon]
   - 2 \alpha\Re Q[\gamma , B_\epsilon] \Big\}\; .
 \end{array}
\end{equation}
Inequalities \eqref{eq:app2} and \eqref{app-est3} imply that there exists a
constant $c_1$ such that for all $\epsilon$ small enough
\begin{equation}\label{eq:bound-1}
 \left| \tr\left((D_Z-m)B_\epsilon\right) \right| < c_1\; .
\end{equation}
Furthermore, using Lemma~\ref{constant-d}, we have
\begin{eqnarray*}
 \tr ( |D_0|^{1/2} |a+\epsilon B_\epsilon|\, |D_0|^{1/2} )
  & = & \tr (|a+\epsilon B_\epsilon|^{1/2} |D_0|
  |a+\epsilon B_\epsilon|^{1/2}) \\
  &\leq &\frac1d
   \tr (|a+\epsilon B_\epsilon|^{1/2} |D_Z|
  |a+\epsilon B_\epsilon|^{1/2}) \\
   & \leq & \frac1d \|
  |D_Z|\,|a+\epsilon B_\epsilon| \|_1 =
  \frac1d \| D_Z (a+\epsilon B_\epsilon) \|_1\, ,
\end{eqnarray*}
which yields, together with \eqref{app-est2.5} and \eqref{app-est3},
$a+\epsilon B_\epsilon\in F$. Thus Lemma \ref{sec:definition-problem} implies
\begin{equation}\label{eq:bound-2}
\begin{split}
  |Q[a + \epsilon B_\epsilon, a + \epsilon B_\epsilon]|& \leq 2 D[ a +
  \epsilon B_\epsilon, a + \epsilon B_\epsilon] \leq \frac{\pi}{2}\|a +
  \epsilon B_\epsilon \|_1
  \tr\left(|D_0| |a + \epsilon B_\epsilon|  \right)\\
  &\leq \frac{\pi}{2d}\|a + \epsilon B_\epsilon \|_1
  \tr\left(|D_Z| |a + \epsilon B_\epsilon|  \right)\\
  & \leq \frac{\pi}{2d}\|a + \epsilon B_\epsilon \|_1 \| D_Z (a + \epsilon
  B_\epsilon) \|_1
\end{split}
\end{equation}
According to \eqref{app-est2.5} and \eqref{app-est3}, we conclude from
\eqref{eq:bound-2} that there exists a constant $c_2$ such that for all
$\epsilon$ small enough
\begin{equation}\label{eq:bound-3}
 \left| Q[a + \epsilon B_\epsilon, a + \epsilon B_\epsilon]\right|
 < c_2\; .
\end{equation}
Now we prove that there exists $c_3$ such that for all $\epsilon$ small enough
\begin{equation}\label{eq:bound-4}
 | Q[\gamma, B_\epsilon]| < c_3 \;.
\end{equation}
We have, for $W^{(B_\epsilon)}$ being the mean field potential associated with
$B_\epsilon\in F$, as defined in \eqref{D-F-potential}
\begin{eqnarray*}
 | Q[\gamma, B_\epsilon]|
 = |\tr (W^{(B_{\epsilon})}\gamma  )|  \;.
\end{eqnarray*}
Moreover
\begin{equation}\label{ppppp}
 \begin{split}
   |\tr(W^{(B_{\epsilon})}\gamma )| &\leq \|W^{(B_{\epsilon})}\gamma\|_1 \leq
   \|W^{(B_{\epsilon})}(D_0)^{-1}\|~\|D_0(D_Z)^{-1} \|~\|D_Z \gamma\|_1 \;.
 \end{split}
\end{equation}
Using Lemmata~\ref{l3prime} and \ref{l3second} with $B_{\epsilon}$ instead of
$\delta$ implies
 \begin{equation}\label{pppppp}
 \|W^{(B_{\epsilon})}(D_0)^{-1}\|\leq 4\|B_\epsilon\|_1\;.
 \end{equation}
 Since $D_0(D_Z)^{-1}$ is bounded, by using Inequalities~\eqref{ppppp},
 \eqref{pppppp} and the fact that $(\Gamma, \Lambda)\in\Upsilon_q$, we obtain
 \eqref{eq:bound-4}.

 Collecting \eqref{eq:bound-1}, \eqref{eq:bound-3}, and \eqref{eq:bound-4}
 yields, together with \eqref{eq:bound-5},
\begin{equation*}
\begin{split}
  \gE(\Gamma, \Lambda_\epsilon) & \!=\! \tr ((D_Z-m)\gamma) + \!\alpha
  Q[\gamma, \gamma] \! - \!  \epsilon \Big\{ \tr((D_Z-m) a )
  + 2\alpha\Re Q[a, \gamma]\Big\} +\mathcal{O}(\epsilon^2)\\
  & = \gE(\Gamma,\Lambda) - \epsilon \tr((D^{(\gamma)}-m) a )
  + \mathcal{O}(\epsilon^2)\\
  & = \gE(\Gamma,\Lambda) - \epsilon \tr(\Lambda^\perp(D^{(\gamma)}-m)\Lambda
  A) - \epsilon \tr(\Lambda(D^{(\gamma)}-m) \Lambda^\perp A^*) +
  \mathcal{O}(\epsilon^2) \;.
\end{split}
\end{equation*}
For $A$ fixed as above, this result remains true for all $\epsilon$ small
enough. Therefore, \eqref{eq:app-crit1} implies
$$
\tr\left(\Lambda^\perp(D^{(\gamma)}-m)\Lambda A\right) +
\tr\left(\Lambda(D^{(\gamma)}-m) \Lambda^\perp A^*\right) = 0 \;.
$$
As at the end of the proof of Theorem \ref{th:hf}, we obtain
$[D^{(\gamma)},\, \Lambda] = 0$.

Finally, exchanging the roles of $\Gamma$ and $\Lambda$ in the above proof
yields $[D^{(\gamma)},\, \Gamma]$.
\end{proof}

\appendix

\section{Some Spectral Properties of Screened
  Coulomb-Dirac Operators\label{section3}} We recall the following result:
\begin{lemma}[Brummelhuis et al \cite{Brummelhuisetal2001}]
  \label{constant-d}
  Let $$
  d:=(1/3)(1-(\alpha Z)^2)^{1/2}((4(\alpha Z)^2+9)^{1/2}-4\alpha Z)$$
  and assume $0\leq \alpha Z <\sqrt{3}/2$. Then
 \begin{equation}\label{constant-d2}
 |D_Z|^2\geq d^2 |D_0|^2\ and\ |D_Z|\geq d |D_0|.
 \end{equation}
\end{lemma}
We would like to compare $|D_0|$ and $|\Dd|$.
\begin{definition}
  Given positive $\alpha$, $Z$, and $m$ and $\delta\in F$, we define
  \begin{equation}
    \label{eq:moduli}
    \cda:=\sup\{c\in\rz\, |\,|\Dd|\geq c|D_0|\}\;.
  \end{equation}
\end{definition}

\begin{lemma}\label{l3prime}
  If $\delta\in F$, then, for all $u\in H^1(\rz^3)\otimes\cz^4$, we have
  $$
  \|\Xd u \| \leq 2 \, \|\delta\|_1 \, \|\nabla u\|\;.
  $$
\end{lemma}
\begin{proof}
  Using the spectral decomposition of $\delta$, we have
\begin{equation*}
  \begin{split}
    \|{\Xd} u \|^2 = & \int\frac{{\delta(x,z)}\, \overline{\delta(x,y)}\,
      u(z)\overline{u(y)}}
    {| \bx -\bz| |\bx -\by|} {\rm d}x\,{\rm d}y\,{\rm d}z\\
    = & \sum_{i,\, j} \lambda_i\, \lambda_j\, \int
    \frac{{\xi_i(x)}\overline{\xi_i(z)} \,\overline{\xi_j(x)}
      {\xi_j(y)}u(z)\overline{u(y)} }
    {| \bx - \bz| | \bx - \by|} {\rm d}x\,{\rm d}y\,{\rm d}z \\
    = & \sum_{i,\, j} \lambda_i \, \lambda_j \, \int{\xi_i(x)} \overline
    \xi_j(x) \left( \int\frac{{\xi_j(y)}\overline{u(y)} }{| \bx - \by|}{\rm
        d}y \int\frac{\overline{\xi_i(z)} u(z)} {| \bx - \bz|}{\rm d}z
    \right) \,{\rm d}x \\
    \leq & \sum_{i,\, j} \int |\lambda_i|\, |\lambda_j|\, |\xi_i(x)|\,
    |\xi_j(x)|\, \bigg[ \left( \int |\xi_j(y)|^2 {\rm d}y \right) \left(
      \int\frac{|u(y)|^2}{| \bx - \by|^2} {\rm d} y
    \right)\bigg] \, {\rm d} x  \\
    \leq & 4 \sum_{i,\, j} \int |\lambda_i|\, |\lambda_j|\, |\xi_i(x)|\,
    |\xi_j(x)|\, \|
    \nabla u \|^2\, {\rm d}x\\
    \leq &4 \, \|\delta\|_1^2 \, \|\nabla u \|^2\;,
\end{split}
\end{equation*}
where we used the Cauchy-Schwarz inequality, the Hardy inequality, and the
identity $\sum_i\,|\lambda_i| = ||\delta||_1 $.
\end{proof}

Similar estimates can be found for the direct part,
\begin{lemma}\label{l3second}
  If $\delta\in F$, then $ \| \fid u\| \leq 2\, \|\delta\|_1 \|\nabla u\| $,
  for all $u\in H^1(\rz^3)\otimes \cz^4$.
\end{lemma}

\begin{proof}
  We have successively
\begin{eqnarray}
        \| \fid u \|^2
& = &  \int |u(y)|^2
     \frac{\delta(x,x)\delta(z,z)}{| \bx - \by|\, | \bz - \by|} {\rm d}x
{\rm d}y {\rm d}z
     \nonumber \\
& = &
      \int\delta(x,x)\delta(z,z) \left(\int\frac{|u(y)|^2}{|
       \bx - \by|\,
     |\bz - \by|} {\rm d}y \right) {\rm d}x {\rm d}z \nonumber \\
 & \leq &
          \frac{1}{2} \int|\delta(x,x)|\, |\delta(z,z)|
         \left(
                \int \frac{|u(y)|^2}{| \bx - \by|^2} {\rm d}y
               +\int \frac{|u(y)|^2}{| \bz - \by|^2} {\rm d}y
        \right)
                {\rm d}x {\rm d}z \nonumber \\
 & \leq &
          4 \int|\delta(x,x)| {\rm d}x \int|\delta(z,z)|
           {\rm d}z\, \|\nabla
u\|^2
     \leq 4\, \|\delta\|_1^{2}\, \|\nabla u\|^2\;,\label{eq:l3second1}
\end{eqnarray}
where we used Hardy's inequality in \eqref{eq:l3second1}.
\end{proof}
A direct consequence of Lemma~\ref{l3second} and the fact that square root is
operator monotone is
\begin{lemma}\label{c:direct}
  If $\delta\in F$, then $|\fid|\leq 2\|\delta\|_1 |D_0|$\;.
\end{lemma}
\begin{lemma}\label{lmodulus}
  If $\delta\in F$ and $\alpha Z \leq \sqrt{3}/2$, the following operator
  inequality holds
  \begin{equation}
    \cda \geq (d-4\alpha \|\delta\|_1)\;. \label{eq:new2}
  \end{equation}
\end{lemma}
\begin{proof}
  This is a direct consequence of Lemmata~\ref{l3prime} and \ref{l3second},
  since we have for all $u$ in $\mathfrak{D}(\Dd)$
  \begin{equation}\label{DdeltaDzero}
    \begin{split}
      \| \Dd u \| & = \|\left(D_Z + \alpha\fid
        -\alpha\Xd\right) u \| \\
      & \geq \| D_Z u\| - \alpha\|\fid u\| - \alpha\| \Xd u\| \\
      & \geq (d - 4 \alpha \|\delta\|_1) \| D_0 u\|\; .
    \end{split}
  \end{equation}
  Therefore $|\Dd|^2 \geq (d - 4 \alpha \|\delta\|_1)^2 |D_0|^2$, which
  concludes the proof since the square root is operator monotone.
\end{proof}

\begin{lemma}\label{l3tierce}
  Assume $\delta\in F$ and $\epsilon>0$, then $\fid, \varphi^{(|\delta|)} \in
  L^{3+\epsilon}(\rz^3)\cap L^\infty(\rz^3)$.
\end{lemma}
\begin{proof}
  As before we denote by $(\lambda_n)_{n\in \mathbb N}$ the eigenvalues of
  $\delta$ and by $(\xi_n)$ a corresponding orthonormal basis of
  eigenfunctions. Since $|\fid|(\bx)| \leq \varphi^{(|\delta|)}(\bx)$, it is
  sufficient to prove the result for $\varphi^{(|\delta|)}$.  \\ {\bf We first
    prove} that $\varphi^{(|\delta|)}\in L^{3+\epsilon}(\rz^3)$. We write
  $\chi_R$ for the characteristic function of the ball with center $0$ and
  radius $R$, and set $\chi_{\bar R}:=1-\chi_R$. We get
  $$
 \begin{array}{ll}
   \||\xi_n|^2 *|\cdot|^{-1}\|_{3+\epsilon} &\leq \|\, |\xi_n|^2
   \|_{\frac32} \, \|\,\chi_R \,|\cdot |^{-1}\, \|_{3/(2+\epsilon)} +
   \| \, \chi_{\bar R}\, |\cdot |^{-1}\,\|_{3+\epsilon}\\
   &\leq \const [\langle \xi_n\,,\, |\bp|\, \xi_n \rangle  + 1]\;,
\end{array}
$$
where we used the Hausdorff-Young inequality and the Sobolev inequality.
Multiplication by $|\lambda_n|$, summation over $n$, and the triangular
inequality yields the result.
$$
\|\varphi^{(|\delta|)}\|_{3+\epsilon} \leq \const \sum_n |\lambda_n|
[\langle \xi_n\,,\, |\nabla|\,\xi_n \rangle + 1] \leq \const \|\delta\|_F\;.
$$
Next we estimate $\|\varphi^{(|\delta|)}\|_\infty$. Using Kato's
inequality, we get $$|\varphi^{(|\delta|)}(\bx)| \leq \sum_n |\lambda_n| \int
dy\, |\xi_n(y)|^2/|\bx-\by|\leq \frac\pi2 \,\sum_n |\lambda_n|\, \langle
\xi_n\,,\,|\bp|\, \xi_n \rangle \leq \frac\pi2\, \|\delta\|_F\;. $$
\end{proof}
\begin{lemma}\label{lem:Xsmallphi}
  If $\delta\in F$, then $|\Xd| \leq \varphi^{(|\delta|)}$.
\end{lemma}
\begin{proof}
  This is a straightforward consequence of the spectral decomposition of
  $\delta$ and the Cauchy-Schwarz inequality.
\end{proof}
\begin{lemma}
  \label{compactness}
  If $\delta\in F$, then $\Wd$ is relatively compact with respect to $D_0$.
\end{lemma}

\begin{proof}
  Since $\fid\in L^4(\rz^3)$ by Lemma \ref{l3tierce}, using an inequality of
  Seiler and Simon \cite[Theorem~4.1]{Simon1979T}, we have
  \begin{equation}\label{eq:estimate-ex-1}
    \|\fid(-\Delta+m^2)^{-1/2} \|_4  \leq
    \|\fid(\cdot)\|_4\,
    \|(|\cdot|^2+m^2)^{-1/2}\|_4 < \infty\;,
  \end{equation}
  implying $D_0^{-1}\fid\in\gS_\infty$.

  We next prove that $\Xd$ is relatively compact with respect to $D_0$. Let us
  denote by $\delta_+$ (respectively $\delta_-$) the positive part
  (respectively the negative part) of $\delta$: $\delta= \delta_+ - \delta_-$,
  $\delta_\pm \geq 0$.

  We have, using $\delta_\pm(x,y) = \overline{\delta_\pm(y,x)}$,
\begin{equation*}
 \begin{split}
   \langle u, (X^{(\delta_\pm)})^2 u\rangle & = \int \overline{u(x)} \left(
     \int \frac{\delta_\pm(x,y)\delta_\pm(y,z)} {|\bx - \by|~|\by -
       \bz|}u(z){\rm d}y{\rm
       d}z\right) {\rm d}x \\
   & \leq \int\frac{h(x)}{h(z)} G(x,z)^\frac12 G(z,x)^\frac12
   \frac{h(z)}{h(x)} {\rm d}x {\rm d}z ,
 \end{split}
\end{equation*}
where $\displaystyle G(x,z):= \left|\int
  \frac{\delta_\pm(x,y)\delta_\pm(y,z)}{|\bx - \by|~|\by - \bz|}{\rm
    d}y\right| |u(x)|^2$ and $h$ is positive and measurable. Applying the
Cauchy-Schwarz inequality and $|\delta_\pm(x,y)|^2\leq
\delta_\pm(x,x)\delta_\pm(y,y)$ yield
 \begin{eqnarray*}
  \langle u, (X^{(\delta_\pm)})^2 u\rangle &\leq& \int
  \frac{h(x)^2}{h(z)^2} \left| \int
  \frac{\delta_\pm(x,y)\delta_\pm(y,z)}{|\bx - \by|~|\by - \bz|}{\rm
  d}y\right||u(x)|^2 {\rm d}x {\rm d}z\\ & \leq & \int
  |u(x)|^2\frac{h(x)^2}{h(z)^2} \left( \int
  \frac{\sqrt{\delta_\pm(x,x)}\delta_\pm(y,y)
  \sqrt{\delta_\pm(z,z)}}{|\bx - \by|~|\by - \bz|}{\rm d}y\right)
  {\rm d}x {\rm d}z.
 \end{eqnarray*}
 Picking $h(x) = \delta_\pm(x,x)^{-1/4}$ implies
 \begin{equation}\label{aleph}
  \begin{split}
    \langle u, (X^{(\delta_\pm)})^2 u\rangle &\leq \int |u(x|^2 \left(\int
      \frac{\delta_\pm (y,y)\delta_\pm(z,z)}{|\bx -\by|~|\by-\bz|}{\rm
        d}y\right) {\rm d}x {\rm d}z\\ & = \int |u(x|^2 \left(\int
      \varphi^{(\delta_\pm)} (\by) \frac{\delta_\pm(y,y)}{|\bx -\by|}{\rm
        d}y\right) {\rm d}x \leq \|\varphi^{(\delta_\pm)}\|_\infty \langle u,
    \varphi^{(\delta_\pm)} u\rangle\ .
  \end{split}
 \end{equation}
 Now, similarly to \eqref{eq:estimate-ex-1}, we have
 $\|\varphi^{(\delta_\pm)}(-\Delta+m^2)^{-1/2} \|_4 <\infty$. Thus, since
 $D_0^{-1}$ is bounded, we obtain $D_0^{-1}\varphi^{(\delta_\pm)}
 D_0^{-1}\in\gS_4(\gH)$. Moreover, by Lemma~\ref{l3tierce} we have
 $\|\varphi^{(\delta_\pm)} \|_\infty < \infty$.  Therefore, using
 \eqref{aleph}, we get $D_0^{-1}(X^{(\delta_\pm)}) ^2 D_0^{-1}\in\gS_4(\gH)$
 and thus $X^{(\delta_\pm)} D_0^{-1}\in\gS_8(\gH)\subset\gS_\infty(\gH)$,
 which implies $\Xd D_0^{-1}$ compact. Using this, and the fact that
 $\varphi^{(\delta)}$ is relatively $D_0$ compact conclude the proof.
\end{proof}
Lemma \ref{compactness}, the Kato-Rellich Theorem, and Weyl's theorem imply
\begin{lemma}\label{lemma10}
  Let $Z >0$ such that $\alpha Z\leq 1$ and let $\delta\in F$. Then the
  operator \break $\Dd = D_Z+\alpha(\fid - \Xd)$ is self-adjoint with domain
  $\mathfrak{D}(\Dd) = \mathfrak{D}(D_Z)$. Moreover
\begin{equation*}
  \sigma_{\rm ess}\left(\Dd\right) = \sigma_{\rm
    ess}\left(D_Z \right) = (-\infty,\, -m]\cup [m,\, +\infty)\;.
\end{equation*}
\end{lemma}

If $\delta$ is a positive density matrix of $q$ electrons with $q < Z+1$, then
the potential of the nucleus prevails giving an attractive Coulomb tail at
infinity. This leads us to expect that $\Dd$ has infinitely many bound states
in the gap accumulating at $m$. The following theorem expresses this
expectation formally.
\begin{theorem}\label{point-spectrum}
  Assume $0\leq\delta\in F$ and $\tr\delta< Z$.  Then the operator $\Dd$ has
  infinitely many eigenvalues in $(0,m)$ accumulating at the point $m$.
\end{theorem}

\begin{proof}
  Pick a function $f\in C_0^\infty(\rz^3)$, normalized in $L^2(\rz^3)$, which
  is also spherically symmetric.  We also define, for all $R>1$, the functions
  $$f_R(x):= R^{-3/2} f(x/R) \mbox{ and } \psi_R = (f_R,0,0,0)\;.$$
  Straightforward calculations using scaling arguments and the specific form
  of $\psi_R$ give
\begin{equation}\label{eq:pp2}
\begin{split}
  \lefteqn{\| \left( D_0 -\frac{\alpha Z}{| \bx |} + \alpha\fid
      -\alpha \Xd\right) \psi_R \|^ 2 }& \\
  &=\, \| \boldsymbol{\alpha}\cdot \frac1i\nabla \psi_R\|^2 + \|m \beta
  \psi_R\|^2 + \| \left( -\frac{\alpha Z}{| \bx |} +
    \alpha\fid - \alpha \Xd\right) \psi_R \|^2 \\
  & \quad + 2\Re \left\langle m\beta \psi_R,\, \boldsymbol{\alpha}\cdot
    \frac1i\nabla \psi_R \right\rangle + 2\Re \left\langle \left(
      -\frac{\alpha Z}{| \bx |} + \alpha\fid - \alpha \Xd\right) \psi_R ,\,
    \boldsymbol{\alpha}\cdot \frac1i\nabla\psi_R
  \right\rangle\\
  &\quad + 2\Re \left\langle m\beta \psi_R,\, \left( -\frac{\alpha Z}{| \bx |}
      + \alpha\fid - \alpha
      \Xd\right)\psi_R\right\rangle\\
  &=\, \|\nabla \psi_R\|^2 + m^2 + \alpha^2\|\left( -\frac{Z}{| \bx |} +\fid-
    \Xd\right)
  \psi_R \|^2 \\
  &\quad + 2 m \alpha\left\langle \psi_R, \left( -\frac{Z}{| \bx |} +
      \fid- \Xd\right)\psi_R\right\rangle\\
  &\leq\, \frac{1}{R^2}\|\nabla f\|^2 + m^2
  +\frac{2\alpha^2}{R^2}\left(\|\frac Z{| \bx |} f\|^2 + 8 (q')^2 \|\nabla
    f\|^2\right)  \\
  & \quad + 2m \left\langle\psi_R, \left( -\frac{\alpha Z}{| \bx |} +
      \alpha\fid- \alpha \Xd\right)\psi_R\right\rangle \\
  &\leq\, m^2 +\frac{c_1}{R^2} + 2m \left\langle\psi_R, \left( -\frac{\alpha
        Z}{| \bx |} + \alpha\fid- \alpha \Xd\right)\psi_R\right\rangle \;,
\end{split}
\end{equation}
where $c_1$ is a constant independent of $R$. Note that we have used
Lemmata~\ref{l3prime} and \ref{l3second} for getting an upper bound of $\|(
\phi^{(\delta)}-X^{(\delta)}) \psi_R\|^2$.

Now, since $f$ is spherically symmetric, using Riesz's rearrangement
inequality and Newton's Theorem, we obtain
\begin{eqnarray*}
        \left\langle \psi_R,\, \fid \psi_R\right\rangle
 & = &
       \int \frac{|f_R(\bx)|^2 \rho_\delta(\by)}{| \bx - \by|}\, {\rm
d}\bx\, \rd \by
       \leq \int \frac{|f_R(\bx)|^2 \rho_\delta^*(\by)}{| \bx - \by|}\,
\rd \bx\,
        \rd \by \\
& \leq &
           \int \frac{|f_R(\bx)|^2 }{| \bx |}\, \rd \bx\int
\rho_\delta^*(\by)\, \rd \by
   \le q'\int \frac{|f_R(\bx)|^2 }{| \bx |}\, \rd \bx \;,
\end{eqnarray*}
where we used, since $\rho_\delta(\by) = \sum_{s=1}^4\delta(y,\,y)$ is
positive, that $$\int \rho_\delta^*(\by)\, \rd \by = \|\rho_{\delta}^* \|_1 =
\|\rho_\delta\|_1 = q'\;.$$
Note also that $\langle \psi_R, \Xd\psi_R\rangle
\geq 0$ and
$$\left\langle \psi_R,\, \frac{1}{| \bx |} \psi_R\right\rangle = \frac1R
\langle f, \frac{1}{| \bx |} f\rangle = \frac{c_2}{R}\;,$$
with $c_2:=\langle
f,\frac{1}{| \bx |} f\rangle>0$.  Thus, we have
\begin{equation}\label{eq:pp3}
\begin{split}
  \left\langle \psi_R,\, \left(-\frac{Z}{| \bx |} +
      \fid-\Xd\right)\psi_R\right\rangle &\leq -\frac{c_2 \,(Z-q')}{R}\; .
\end{split}
\end{equation}
Now \eqref{eq:pp2} and \eqref{eq:pp3} imply
\begin{equation*}
\begin{array}{ll}
   \|\left[ \Dd - E\right]\psi_R\|^2
  & = \|  \Dd\psi_R\|^2 + E^2  -2E \left\langle
      \psi_R, \Dd
      \psi_R \right\rangle \\
   &\leq m^2 +\frac{c_1}{R^2} + 2m\langle\psi_R,\alpha\left(
   -\frac{Z}{|\bx|} + \fid - \Xd\right)\psi_R\rangle +E^2 \\
   &\quad -2E \langle\psi_R,\alpha\left(
   -\frac{Z}{|\bx|} + \fid - \Xd\right)\psi_R\rangle -2 E
   \langle\psi_R , D_0 \psi_R \rangle \\
   & \leq m^2 +\frac{c_1}{R^2} + 2(m-E)\langle\psi_R,\alpha\left(
   -\frac{Z}{|\bx|} + \fid - \Xd\right)\psi_R\rangle\\
 &\quad +E^2 + \frac{c_1}{R^2}
   -2 m E \\
  & = (m-E)^2 + 2\frac{c_1}{R^2} - 2(m-E)\alpha\frac{c_2(Z-q')}{R} \;.
\end{array}
\end{equation*}
Therefore, if $R$ is large enough, we get the inequality
$$
\| \left( \Dd - E \right) \psi_R \| < |m-E| \;,
$$
which implies, by taking $E=m/2$, that $ \Dd $ has at least one eigenvalue
$\lambda_1$ in $(0,m)$.

Now by taking $E=(m+\lambda_1)/2$, by the same argument as above, one gets a
second eigenvalue $\lambda_2\in(\lambda_1,m)$. The iteration of this procedure
yields an infinite sequence of eigenvalues $(\lambda_n)$ of $\Dd$ in $(0,m)$
tending to $m$.
\end{proof}
A similar result holds for the no-pair Dirac-Fock operator.
\begin{lemma}\label{gs}
  Assume that $\delta$ and $\gamma$ are two positive definite finite rank
  density matrices. Assume, in addition, that $\gamma$ is purely electronic
  having particle number not exceeding $Z$, i.e., we have
  $$\delta\in F\,,\,\delta\geq 0\,,\,\gamma\in F\cap
  S^{\delta}\,,\,\Ld_+\gamma\Ld_+ = \gamma\mbox{ and } \tr\gamma < Z\;.
  $$
  Moreover we assume $$
  d-2\alpha(2\tr\delta +\tr\gamma)\geq 0\;.$$
  Then
  the no-pair Dirac-Fock Hamiltonian $\Ld_+D^{(\gamma)} \Ld_+$ has infinitely
  many eigenvalues in $(0, m)$.
\end{lemma}\begin{proof}
  We first prove that
 \begin{equation}\label{eq:locate-proj-DF}
  \sigma_{\mathrm{ess}}(\Ld_+ D^{(\gamma)} \Ld_+) =
  [m,+\infty)
 \end{equation}
 Lemma~\ref{compactness} implies that $W^{(\delta)}D_0^{-1}$ and
 $W^{(\gamma)}D_0^{-1}$ are compact. By assumption, we have $(d-4\alpha
 \|\delta\|_1) >0$; thus \eqref{DdeltaDzero} implies that $D_0 (\Dd)^{-1}$ is
 bounded. This yields
 $$
 (W^{(\gamma)}-W^{(\delta)})(\Dd)^{-1} = W^{(\gamma)} D_0^{-1} D_0
 (\Dd)^{-1} \!- W^{(\delta)} D_0^{-1} D_0 (\Dd)^{-1} \!\in\gS_\infty(\gH)\; ,
 $$
 and
 $$
 \Ld_+ (W^{(\gamma)}-W^{(\delta)}) \Ld_+ (\Ld_+ \Dd
 \Ld_+)^{-1}\in\gS_\infty(\gH)\; .
 $$
 Using the Kato-Rellich Theorem, this implies
 $$
 \sigma_{\mathrm{ess}}( \Ld_+ D^{(\gamma)}\Ld_+ ) =
 \sigma_{\mathrm{ess}}(\Ld_+(\Dd + W^{(\gamma)}-W^{(\delta)})\Ld_+) =
 \sigma_{\mathrm{ess}}(\Ld_+ \Dd \Ld_+)\; .
 $$
 Together with Lemma~\ref{lemma10}, this proves \eqref{eq:locate-proj-DF}.

 Set $q':=\tr\delta$ and $q:=\tr\gamma$. We denote by $\varphi^{(\gamma)}$ and
 $X^{(\gamma)}$ respectively the direct and exchange operators associated to
 $\gamma$, defined by replacing $\delta$ with $\gamma$ in \eqref{direct} and
 \eqref{exchange}. For all $\displaystyle u\in \Ld_-
 \left(H^1(\rz^3)\otimes\cz^4\right)$, we have
\begin{eqnarray}
 { \langle u,\,  D^{(\gamma)}  u\rangle }
 & = &  \langle u,  \Dd  u\rangle +
  \alpha \langle u,  \varphi^{(\gamma)}  u\rangle
  - \alpha\langle u, X^{(\gamma)}  u\rangle
  -  \alpha\langle u, \fid  u\rangle
  +  \alpha\langle u,  \Xd  u\rangle\nonumber \\
 & \leq & - \langle u, |\Dd |u\rangle + \alpha\langle u,
 \varphi^{(\gamma)} u\rangle\;,
 \label{eq:use}
\end{eqnarray}
where, in Inequality \eqref{eq:use}, we used Lemma \ref{lem:Xsmallphi} and the
fact that $X^{(\gamma)} \geq 0$.  Now, from Lemmata~\ref{c:direct} and
\ref{lmodulus} we obtain
$$
{ \langle u,\, D^{(\gamma)} u\rangle } \leq \left( -( d - 4\alpha q') +
  2\alpha q\right) \langle u,\, |D_0| u\rangle \;.
$$
Since $ -( d - 4\alpha q') + 2\alpha q<0 $, we have
\begin{equation}\label{diag-dirac-neg}
\Ld_- D^{(\gamma)} \Ld_- \leq 0\;.
\end{equation}
Now, thanks to \eqref{diag-dirac-neg}, one can apply \cite[Theorem
3]{GriesemerSiedentop1999}. With their notations, since ${\mathfrak D}(\Dd) =
{\mathfrak D} (D^{(\gamma)})$, we first define $\displaystyle{\mathfrak Q}_\pm
:= \left({\mathfrak D} (D^{(\gamma)})\right)\cap \Lambda_\pm^{(\delta)} \gH$.
Then, denoting by $\mu_n(A)$ the $n^{\rm th}$ lowest eigenvalue of the
operator $A$, we obtain
\begin{eqnarray*}
  m> \mu_n\left( D^{(\gamma)} |_{{\Lambda_+}^{(\gamma)}\gH} \right) &
\geq &
  \inf_{\buildrel{M_+\subset {\mathfrak Q_+}}\over{\dim(M_+) =n}}
 \sup_{\buildrel{\zeta\in M_+\oplus {\mathfrak Q_-}}\over{\|\zeta\|
=1}}
 \langle \zeta,\, D^{(\gamma)}\zeta \rangle \;.
\end{eqnarray*}
Moreover,
\[
\inf_{\buildrel{M_+\subset {\mathfrak Q_+}}\over{\dim(M_+) =n}}
\sup_{\buildrel{\zeta\in M_+\oplus {\mathfrak Q_-}}\over{\|\zeta\| =1}}
\langle \zeta,\, D^{(\gamma)}\zeta\rangle\geq \inf_{\buildrel{M_+\subset
    {\mathfrak Q_+}}\over{\dim(M_+) =n}} \sup_{\buildrel{\zeta\in
    M_+}\over{\|\zeta\| =1}} \langle \zeta,\, D^{(\gamma)} \zeta\rangle =
\mu_n\left( \Ld_+ D^{(\gamma)} \Ld_+ \right)\;.
\]
Therefore $$
m> \mu_n \left( D^{(\gamma)} |_{\Lambda_+^{(\gamma)}\gH} \right)
\geq \mu_n\left( \Ld_+ D^{(\gamma)} \Ld_+ \right)\;. $$
Since from
Theorem~\ref{point-spectrum} the operator $D^{(\gamma)}$ has infinitely many
eigenvalues in $(0,\, m)$, using \eqref{eq:locate-proj-DF}, we finally get the
expected result.
\end{proof}

If we have a non-negative spherical symmetric density $\rho$ with
$$
q:=\int \rho(\by) d\by \leq Z\;,$$
then $$-Z/|\cdot|+\rho*
\,|\cdot|^{-1}\geq 0\;.
$$
This implies that the $n$-th eigenvalue of $D_Z+\varphi$ can be estimated
from below by the $n$-th eigenvalue of $D_{Z-q}$. For a non-spherical
symmetric potential this situation for the positive eigenvalues is disturbed
only slightly.

\begin{lemma}
  \label{sec:some-spectr-prop-1}
  Assume $0\leq\delta\in F$.  Then $\chi_{(0,m)}(\Dd)(\Dd-m)$ is a
  Hilbert-Schmidt operator.
\end{lemma}
\begin{proof}
  Since $\delta\in F$, by writing $W^{(\delta)}D_Z^{-1} = W^{(\delta)}D_0^{-1}
  D_0 D_z^{-1}$, Lemmata~\ref{constant-d} and \ref{compactness} imply that
  $W^{(\delta)}$ is relatively compact with respect to $D_Z$. Moreover, since
  $\delta\geq 0$, we have $W^{(\delta)}\geq 0$ (see
  lemma~\ref{lem:Xsmallphi}).


  Let $\lambda_0(0)\leq \lambda_1(0) \leq \ldots$ be the ordered positive
  eigenvalues of $D_Z$, including multiplicity. We first prove that for all
  $\epsilon\in [0,1]$, there exist $N_+(\epsilon)$ and $N_-(\epsilon)$ in
  $\{-\infty\}\cup\mathbb{Z}\cup\{+\infty\}$, with $N_-(\epsilon) \leq
  N_+(\epsilon)$, and $\displaystyle\left( \lambda_k(\epsilon)\right)_{k\in \{
    N_-(\epsilon),\ldots, N_+(\epsilon) \} }$ such that

  i) $\sigma\left( D_Z + \epsilon\alpha\Wd \right)\cap (-m,m) = \{
  \lambda_{N_-(\epsilon)}(\epsilon), \lambda_{N_-(\epsilon)+1}(\epsilon),
  \ldots, \lambda_{N_+(\epsilon)}(\epsilon)\}$

  ii) $\lambda_{N_-(\epsilon)}(\epsilon) \leq
  \lambda_{N_-(\epsilon)+1}(\epsilon)\leq \ldots \leq
  \lambda_{N_+(\epsilon)}(\epsilon)$

  iii) For all $k\in \{N_-(\epsilon), \ldots, N_+(\epsilon)\}$ the functions
  $\lambda_k(\epsilon)$ are continuous, monotone increasing.

  Using Kato's perturbation theory for isolated eigenvalues and numbering the
  eigenvalues that are crossing with respect to their ordering -- namely the
  largest after crossing gets the highest index --, yields i) and ii).
  Continuity is also a consequence of perturbation theory for eigenvalues. The
  asserted monotonicity is a consequence of positivity of $\Wd$.

  We next show that the number $M$ of eigenvalues $\lambda_k(\epsilon)$ that
  crosses zero when $\epsilon$ increases from $0$ to $1$ is finite. By the
  Birman-Schwinger principle,
  $$
  M =\#\{\lambda\in[1,\infty)\,| \, \lambda\ \mathrm{ eigenvalue\ of}\
  R\}\;,
  $$
  where $R=-(\Wd)^{1/2}D_Z^{-1}(\Wd)^{1/2}$ is the Birman-Schwinger kernel
  with energy $0$. Thus, since we count only the eigenvalues of $R$ larger
  than $1$, we have, using the notation $W:=\Wd$ and $\varphi:=\fid$,
\begin{equation}\label{eq:bs-1}
\begin{split}
  M & \leq \|R\|_4^4 \leq \|W^{\frac12}\varphi^{-\frac12}\varphi^{\frac12}
  |D_0|^{-\frac12} |D_0|^{\frac12} D_Z^{-1} |D_0|^{\frac12} |D_0|^{-\frac12}
  \varphi^{\frac12} \varphi^{-\frac12} W^{\frac12}\|_4^4\\
  & \leq \|W^{\frac12} \varphi^{-\frac12}\|^8 \; \|\, |D_0|^{\frac12}
  D_Z^{-\frac12}\|^8\; \|\varphi^{\frac12} |D_0|^{-\frac12}\|_8^8
\end{split}
\end{equation}
Since $0\leq \Wd = \fid -\Xd \leq \fid$, we get
\begin{equation}\label{eq:bs-2}
 \| (\Wd)^{1/2} (\fid)^{-1/2} \| \leq 1 .
\end{equation}
Lemma~\ref{constant-d} yields
\begin{equation}\label{eq:bs-3}
 \| |D_0|^{1/2}D_Z^{-1/2}\| \leq d^{-1} <\infty .
\end{equation}
Using \eqref{constant-d2} and \cite[Theorem~4.1]{Simon1979T} as for Inequality
\eqref{eq:estimate-ex-1} in the proof of Lemma~\ref{compactness} yields
\begin{equation}\label{eq:bs-4}
 \| (\fid)^{1/2} |D_0|^{-1/2} \|_8 \leq \|\fid(\cdot)\|_4 ^{1/2}
 \| \sqrt{|\cdot|^2 + m^2}\|_4^{1/2} <\infty .
\end{equation}
Collecting \eqref{eq:bs-1}-\eqref{eq:bs-4} proves $M<\infty$.

Therefore, apart from the eigenvalues $\lambda_0(1), \lambda_1(1),\ldots $,
the operator $\Dd$ has only finitely many other eigenvalues in $[0,m]$. Thus,
$\chi_{(0,m)}(\Dd)(\Dd-m)\in\gS_2(\gH)$ follows if the series $\sum_{k\geq 0}
(\lambda_k(1)-m)^2\leq\sum_{k\geq 0} (\lambda_k(0)-m)^2$ is convergent. At
this point we remind the reader that the relativistic hydrogen eigenvalues
$\lambda_k(0)$ can be grouped into ``multiplets'' of $2n^2$ eigenvalues
corresponding to one non-relativistic eigenvalue $m-Z^2\alpha^2/n^2$. Each
element of such a multiplet can be bounded from below by the previous
non-relativistic eigenvalue $m-Z^2\alpha^2/(n-1)^2$. Thus, up to an
unessential multiplicative constant, $\sum_{n\geq 2} n^2/(n-1)^4$ is a
convergent majorant.  This proves the claim.
\end{proof}

\end{document}